\documentclass{amsart}

\usepackage{amsmath}
\usepackage{amsthm}
\usepackage{latexsym}
\usepackage{amsfonts}
\usepackage{amssymb}
\usepackage{bm,dsfont}
\usepackage{color}
\usepackage{graphicx}
\usepackage{enumerate}
\usepackage{subfigure}

\usepackage[normalem]{ulem}
\usepackage{mathrsfs}
\usepackage{mathtools}

\usepackage{comment}

\newtheorem{proposition}{Proposition}
\newtheorem{theorem}{Theorem}

\theoremstyle{definition}

\definecolor{darkgreen}{rgb}{0,0.4,0.3}
\definecolor{gray}{rgb}{0.5,0.5,0.5}

\newcommand{\R}{\mathbb{R}} 
\newcommand{\real}{\mathbb R} 
\newcommand{\C}{\mathbb C}

\newcommand{\naturale}{\mathbb N} 
\newcommand{\half}{\tfrac{1}{2}} 
\newcommand{\mo}[1]{\left| #1 \right|} 
\newcommand{\floor}[1]{\left\lfloor{#1}\right\rfloor}
\newcommand{\ceil}[1]{\left\lceil{#1}\right\rceil}

\newcommand{\dual}[2]{\left\langle\,#1\,,\,#2\,\right\rangle} 

\newcommand{\hh}{\mathcal{H}} 
\newcommand{\lh}{\mathcal{L(H)}} 
\newcommand{\lc}{\mathcal{L}(\C^2)} 
\newcommand{\kb}[2]{|#1\rangle\langle#2|} 
\newcommand{\no}[1]{\left\|#1\right\|} 
\newcommand{\tr}[1]{{\rm tr}\left[#1\right]} 
\newcommand{\id}{\mathds{1}} 
\newcommand{\rank}[1]{{\rm rank}(#1)} 

\newcommand{\va}{{\bm a}} 
\newcommand{\vb}{{\bm b}} 
\newcommand{\ve}{{\bm e}} 
\newcommand{\vm}{{\bm m}} 
\newcommand{\vmh}{{\bm{\hat{m}}}} 
\newcommand{\vr}{{\bm r}} 
\newcommand{\vt}{{\bm t}} 
\newcommand{\vx}{{\bm x}} 
\newcommand{\vsigma}{{\bm \sigma}}

\newcommand{\M}{\mathsf{M}}
\newcommand{\N}{\mathsf{N}}
\newcommand{\Oo}{\mathsf{O}}

\newcommand{\en}{\mathcal{E}} 
\newcommand{\enf}{\mathcal{F}} 
\newcommand{\Pg}{\mathbf{P}} 
\newcommand{\Eg}{\mathbf{E}} 
\newcommand{\Epost}{\mathbf{E}^{\mathrm{post}}}
\newcommand{\Epre}{\mathbf{E}^{\mathrm{prior}}}
\newcommand{\Ef}{\mathbf{E}_{f}} 
\newcommand{\Efpost}{\mathbf{E}_{f,\alpha}^{\mathrm{post}}} 
\newcommand{\Efpre}{\mathbf{E}_{f,\alpha}^{\mathrm{prior}}} 

\begin{document}

\title[]{Quantum guessing games with posterior information}

\author{Claudio Carmeli}

\author{Teiko Heinosaari}

\author{Alessandro Toigo}

\begin{abstract}
Quantum guessing games form a versatile framework for studying different tasks of information processing. A quantum guessing game with posterior information uses quantum systems to encode messages and classical communication to give partial information after a quantum measurement has been performed. 
We present a general framework for quantum guessing games with posterior information and derive structure and reduction theorems that enable to analyze any such game. 
We formalize symmetry of guessing games and characterize the optimal measurements in cases where the symmetry is related to an irreducible representation. 
The application of guessing games to incompatibility detection is reviewed and clarified.
{All the presented main concepts and results are demonstrated with examples.}
\end{abstract}

\maketitle


\section{Introduction}\label{sec:intro}

Information processing, both classical and quantum, is ultimately about getting a desired output from a given input. This can be seen as a guessing game, where the aim is formalized as a score function that gives high scores for successful outputs and no scores for unsuccessful outputs. 
The guessing game setup is a natural translation of many different information processing scenarios and it is therefore a useful framework for studying the advantages that manipulation of quantum systems can give in information processing tasks.
The guessing game can be a communication scenario, where Alice tries to transmit information to Bob, possibly simultaneously hiding it from others.
Or it can be a computing scenario, where Alice chooses an input string and then runs a computation on it (in this case Alice and Bob can be the same person).
Our interest is in quantum guessing games, where the transmitted information is encoded into quantum states and then decoded by a quantum measurement. There can be processing between encoding and decoding, but this can all be seen as a part of the measurement since we put no restrictions on it.

In both of the previously mentioned scenarios it is possible that Alice, or someone else, sends partial information after Bob has already performed a measurement. 
In the presently investigated scenario this later sent information is classical and we call these games \emph{(quantum) guessing games with posterior information}.
In the computing scenario this kind of game can be seen as a hybrid computation, where one runs classical and quantum computing in parallel and uses both to conclude the final result.
The classical part of a computation may, for example, find one instance that is known to be incorrect with certainty while the quantum part tries to find the correct answer even if some error is expected. 
The final guess takes into account both parts and is then typically better than each of them alone. 

The main aim of this paper is to present a clear framework for different types of guessing games with posterior information.
We show that any such game can be written in a certain kind of standard form and, further, the calculation of the maximal average score in a given game reduces to the calculation of the usual discrimination success probability of a so-called auxiliary state ensemble.
We formulate symmetry for guessing games with posterior information and present the solution of a symmetric scenario when the symmetry is related to an irreducible representation.
With examples we demonstrate that it is, indeed, possible to calculate the best average score analytically in many interesting cases. 
In our exemplary cases we derive the solutions for a class of encodings in a qubit system (the angle between the states of the encodings being a free parameter) and this therefore enables to make comparisons and observations that a bunch of numerical solutions could not provide. 

It is instructive to compare guessing games with posterior information to similar scenarios where the classical partial information is given to Bob before he is performing a measurement.
We call this kind of scenario a \emph{(quantum) guessing game with prior information}.
Typically prior information allows Bob to adjust and optimize his measurement in a more clever way than when the same partial information is given afterwards.
This difference in average scores is the basis of a method that uses guessing games in the detection of quantum incompatibility.  
We reformulate the incompatibility detection method in the present general framework, recall the known results and point out some open questions. 
We further characterize a class of encodings for which prior and posterior information are equally valuable.
For these encodings the timing of partial information is therefore irrelevant. 
The fact that in quantum guessing games the timing of partial information can change the maximal average score is the essential difference to classical guessing games. 
This observation may aid in finding new applications of quantum guessing games where the manipulation of quantum systems boosts information processing.  

At this point, it is in order to briefly comment on related scenarios that have been investigated earlier. 
As already mentioned, in this work by a guessing game we mean a task that one party (Alice) sets for another party (Bob) and where the goal is specified by a score function. The aim of Bob is to maximize the average score and the basic question is how to do it, i.e., what are the optimal actions and what is the maximal average score that can be achieved with those actions. By calling this scenario a guessing game we want to make a distinction with more general scenarios and emphasize their different basic questions and tasks. Nevertheless, one should note that the terminology varies and a guessing game can mean something different in other contexts. Also, the resources that are available for the players vary in different investigations. 
For example, a \emph{prepare-and-measure scenario} refers to a similar setting, although it is often assumed that Alice and Bob get independent inputs and that they might also have either shared randomness or shared entanglement. The basic question is to characterize the correlations that Alice and Bob can generate  and to derive conclusions on properties (e.g. dimension) of the system and devices \cite{GaBrHaAc10,TaKaVeRoBr18,PoBrNeScCh20}.
A special type of these games are \emph{non-local games}, which usually mean games with several space-separated players that can communicate only with a referee but can have a preliminarily agreed joint strategy \cite{BrBrTa05,Toner09,AmKrNaRi10}. 
The basic question is to see the effect of shared entanglement or other quantum resources. 
Another related topic is that of \emph{input-output processes} and their analysis. 
This is a broad topic and has diverse research questions, e.g. to characterize some properties of the intermediate quantum dynamics \cite{CaHeMiTo19JMP,UoKrAb19,Mori20}.
All the previously mentioned scenarios are under active research and their applications grow rapidly.
A general conclusion is that by analyzing how the use of certain resources can facilitate the achievement of specific tasks has proven to be a powerful way to clarify fundamental aspects of quantum theory.

The benefit of restricting the current work to (later precisely specified) guessing games with posterior classical information is that we can present an in-depth analysis and derive a reduction theorem for all such games.
Our investigation is organized as follows. 
In Section \ref{sec:guessing} we recall the basics of usual state discrimination and, more generally, guessing games with arbitrary score function. 
This scenario is expanded in Section \ref{sec:posterior} to cover guessing games with posterior information, which are the focus of the current work.
These games can be recast in the so-called standard form, explained in Section \ref{sec:reduction}. 
Strikingly, the maximal average score in any guessing game with posterior information equals with the maximal success probability in the usual state discrimination game of a related auxiliary state ensemble. 
This simple but important result is also treated in Section \ref{sec:reduction} and it implies that all known methods to solve state discrimination games are applicable in our more general setting. 
Section \ref{sec:prior} reviews the connection of guessing games to incompatibility detection. 
In Section \ref{sec:symmetry} we formulate symmetry of guessing games with posterior information and show how it can be used to calculate the maximal average score in symmetric scenarios.
Three different kind of examples that demonstrate all the presented main concepts and results are treated in Section \ref{sec:qubit}.
Finally, in Section \ref{sec:conc} we summarize our conclusions and point out some new directions for future investigations.

\section{Guessing games}\label{sec:guessing}

\subsection{State discrimination}\label{sec:state}

We will deal with finite dimensional quantum systems and measurements with a finite number of outcomes.  
We fix a $d$-dimensional, complex Hilbert space $\hh$, denote by $\lh$ the set of all its linear operators and say that $\varrho$ is {\em state} on $\hh$ if it is a positive element of $\lh$ (i.e.~$\varrho$ is selfadjoint with nonnegative eigenvalues) and $\tr{\rho} = 1$. We denote by $|X|$ the cardinality of a finite set $X$. A {\em measurement} on $\hh$ with the outcome set $X$ is a map $\M:X\to\lh$ such that $\M(x)$ is positive for all $x$ and $\sum_x\M(x)=\id$. 
A {\em state ensemble} on $\hh$ with the label set $X$ is a map $\en:X\to\lh$ such that $\en(x)$ is positive for all $x$ and $\sum_x \tr{\en(x)}=1$. Any state ensemble can be written as a product $\en(x)=p(x)\,\varrho_x$, where $(\varrho_x)_{x\in X}$ is a family of states on $\hh$ and $p:x\mapsto\tr{\en(x)}$ is a probability distribution on $X$.

In the usual {\em minimum error state discrimination}, the system is prepared in one of several possible states $\varrho_x$, $x\in X$, and the task is to guess the correct state by performing a measurement. This can be seen as a scenario where two parties communicate by one of them sending one classical message $x$ -- the label of the state -- to the other, and to this aim he encodes $x$ into a quantum system. The encoding is then described by a state ensemble $\en$, in which the probability distribution $p$ is the prior probability of labels to occur and $x\mapsto \varrho_x$ is the actual encoding. For any measurement $\M$ with the outcome set $X$, we denote by $\Pg(\en;\M)$ the \emph{guessing probability}, given as
\begin{equation}\label{eq:guessing_prob}
\Pg(\en;\M) =  \sum_{x} \tr{\en(x)\,\M(x)} =  \sum_{x} p(x)\, \tr{\varrho_x\,\M(x)} \, .
\end{equation}
The maximal guessing probability for $\en$ is denoted as
\begin{equation}
\Pg(\en) =  \max_{\M}\,  \Pg(\mathcal{E};\M) \, ,
\end{equation}
where the optimization is over all measurements with the outcome set $X$.
We refer to reviews \cite{BaCr09,Bae13,BaKw15} for more background and details on state discrimination.

There is a communication task which is, in a sense, opposite to state discrimination and therefore called \emph{antidiscrimination}, also antidistiguishability or state exclusion. 
As in state discrimination,  the system is prepared in one of several possible states $\varrho_x$, $x\in X$.
But now the task is to guess one of the labels different from the encoded label $x$. 
Hence, the success probability (i.e. the probability of guessing a label different from the encoded label) in the antidiscrimination task is $1-\Pg(\en;\M)$, and its optimization amounts to minimizing -- instead of maximizing -- the guessing probability \eqref{eq:guessing_prob}.
Apart from its simplicity, antidiscrimination has turned out to be a fruitful notion.
For instance, it has played a key role in discussions on the controversy between epistemic and ontic interpretations of quantum states \cite{PuBaRu12,Hardy13,Leifer14}, while a connection to quantifications of quantum resources has been revealed in \cite{UoBuKrPeBr20} 

\subsection{General form of guessing games}\label{sec:game}

\begin{figure}[h!]
\centering
\includegraphics[scale=0.7]{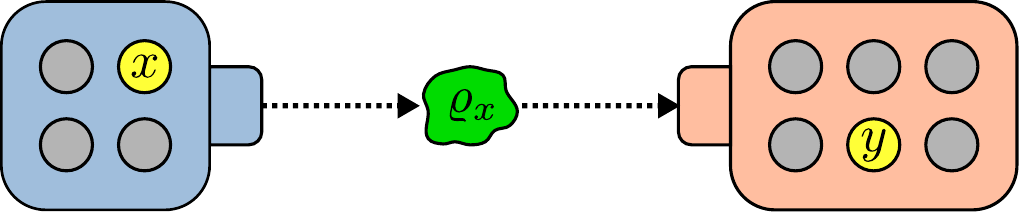}
\caption{In a guessing game one party (Alice) encodes a classical message $x$ into a quantum state, and then she sends the state to another party (Bob). Bob then performs a measurement and thus he obtains an outcome $y$. {By comparing $x$ to $y$, a score $f(x,y)$ is assigned to the game.} 
 The set $X$ of possible choices for the sent message and the set $Y$ of possible outcomes for the performed measurement need not coincide.
\label{fig:game}}
\end{figure}

A \emph{guessing game} can be something different than discrimination or antidiscrimination, although the basic idea is the same (see Figure \ref{fig:game}).
Generally, we have a \emph{score function} $f:X\times Y \to [0,1]$ and the associated \emph{average score} is given as
\begin{equation}\label{eq:Ef}
\Ef(\en;\M) =  \sum_{x,y} f(x,y)\, \tr{\en(x)\,\M(y)} = \sum_{x,y} f(x,y)\,p(x)\, \tr{\varrho_x\,\M(x)} \, .
\end{equation}
The input and output label sets $X$ and $Y$ can be different (some examples are presented shortly).
We are often considering a scenario where a state ensemble $\en$ is given and the used measurement $\M$ is optimized to give as high average score as possible. 
The maximal average score is denoted by
\begin{equation}\label{eq:def_E(E)}
\Ef(\en) =  \max_{\M}\,  \Ef(\en;\M) \, .
\end{equation}
In a typical guessing game some pairs $(x,y)\in X\times Y$ are wanted (successful guess) and other pairs are unwanted (unsuccessful guess). 
If we assign values $f(x,y)=1$ for wanted pairs and $f(x,y)=0$ for unwanted pairs, then the average score  $\Ef(\en;\M)$ equals with the probability of getting a wanted pair.
Intermediate scores (i.e. $0<f(x,y)<1$) are also possible and can be e.g.~used to give some reward if the guess is almost wanted but not exactly. 

In the previously discussed state discrimination we set $X=Y$ and choose a score function $f$ which assigns nonzero values to all elements on the diagonal of $X\times X$ and $f(x,y) = 0$ for all $x\neq y$. 
If we additionally require that $f$ takes only values $0$ and $1$, then we get the standard discrimination score function $f(x,y)=\delta_{x,y}=:f_\delta(x,y)$, for which $\Eg_{f_\delta} = \Pg$.
In the antidiscrimination task one aims to get any other outcome than the sent message $x$, thus we choose a score function $f$ such that $f(x,x)=0$ and $f(x,y)>0$ for $y\neq x$. 
If we further require that $f$ takes only values $0$ and $1$, then we obtain the standard antidiscrimination score function $f=1-f_\delta$.

The general formulation of guessing games directs us to see that there are natural generalizations of discrimination and antidiscrimination games to the cases in which the receiver is allowed to guess several (fixed integer $2\leq k< |X|$) outcomes instead of one.  
To formulate these type of guessing games, we choose $Y=\{S\subset X : \mo{S} = k\}$ and $f$ such that $f(x,S)=1$ for $x\in S$ and $f(x,S)=0$ otherwise. 
The receiver hence gets a score if and only if the input $x$ is contained in the guessed set $S$.
In the respective generalization for antidiscrimination games we choose $f$ such that $f(x,S)=1$ for $x\notin S$ and $f(x,S)=0$ otherwise.

\subsection{Reduction to usual state discrimination}\label{sec:red}

Different score functions determine different kind of guessing games and they can have quite diverse goals. 
However, the calculation of the maximal average score and determination of optimal measurement strategy are similar in all guessing games. 
In fact, following \cite[Section 2.2.2]{SSQT01}, any guessing game can be recast as a discrimination game by suitably redefining the state ensemble at hand. 
To this aim, we set
\begin{equation}\label{eq:Delta}
\Delta(\en,f) = \sum_{x,y} f(x,y) \, p(x)
\end{equation}
and whenever this constant is nonzero we further define the \emph{auxiliary state ensemble} $\en_f$ with the label set $Y$ as
\begin{equation}\label{eq:enf_0}
\en_f (y) = \Delta(\en,f)^{-1} \sum_x f(x,y)\,\en(x) \,.
\end{equation}
With this definition we have the equalities
\begin{gather}
\Ef(\en;\M) = \Delta(\en,f)\ \Pg(\en_f;\M) \,, \label{eq:aux_Pgfpost_1} \\
\Ef(\en) = \Delta(\en,f)\ \Pg(\en_f)\,. \label{eq:aux_Pgfpost_2}
\end{gather}
In this way a guessing game with an arbitrary score function $f$ is recast in a usual state discrimination game for the respective auxiliary state ensemble.

We remark that the precondition $\Delta(\en,f)\neq 0$ mentioned earlier means that $f(x,y)\neq 0$ for some $x,y$ with $\en(x)\neq 0$.
If $\Delta(\en,f)=0$, the auxiliary state ensemble can be defined in an arbitrary way without changing \eqref{eq:aux_Pgfpost_1}-\eqref{eq:aux_Pgfpost_2}, since in that case $\Ef(\en;\M) = 0$ for all $\M$ and thus \eqref{eq:aux_Pgfpost_1}-\eqref{eq:aux_Pgfpost_2} are satisfied for any choice of $\en_f$. 

We end this section with an upper bound for $\Pg(\en)$ which despite its simplicity will be quite useful in the later developments (Sections \ref{sec:symmetry} and \ref{sec:qubit}).
It has the same derivation as \cite[Proposition 2]{CaHeTo18}.

\begin{proposition}\label{prop:Pbound}
For a state ensemble $\en$ with the label set $X$, we denote by $\Lambda(\en)$ the largest eigenvalue of all the operators $\en(x)$, $x\in X$. 
Then,
\begin{equation}\label{eq:Pbound}
\Pg(\en)\leq d \, \Lambda(\en) \,.
\end{equation}
The above equality is attained if and only if there exists a measurement $\M$ with the outcome set $X$ satisfying $\en(x)\,\M(x) = \Lambda(\en)\,\M(x)$ for all $x\in X$. If this is the case, then $\Pg(\en) = \Pg(\en;\M)$ for such a measurement.
\end{proposition}

\begin{proof}
If $\lambda(x)$ is the largest eigenvalue of the operator $\en(x)$, we have $\lambda(x)\,\id-\en(x) \geq 0$, and then
\begin{align*}
& \lambda(x)\,\tr{\M(x)}-\tr{\en(x)\,\M(x)} = \tr{\big(\lambda(x)\,\id-\en(x)\big)\M(x)} \\
& \qquad\quad = {\rm tr}\Big\{\Big[\big(\lambda(x)\,\id-\en(x)\big)^{\frac{1}{2}}\M(x)^{\frac{1}{2}}\Big]^* \Big[\big(\lambda(x)\,\id-\en(x)\big)^{\frac{1}{2}} \M(x)^{\frac{1}{2}}\Big]\Big\} \geq 0\,.
\end{align*}
In this expression, the last equality is attained if and only if $\big(\lambda(x)\,\id-\en(x)\big)\M(x) = 0$, that is, $\en(x)\,\M(x) = \lambda(x)\,\M(x)$. It follows that
\begin{align*}
\Pg(\en;\M) & = \sum_x \tr{\en(x)\,\M(x)} \leq \sum_x \lambda(x)\,\tr{\M(x)} \leq \sum_x \Lambda(\en)\,\tr{\M(x)} = \Lambda(\en)\,\tr{\id} \\
& = d \, \Lambda(\en) \,,
\end{align*}
where all the equalities are attained if and only if $\en(x)\,\M(x) = \lambda(x)\,\M(x)$ for all $x$ and $\M(x) = 0$ for all $x$ such that $\lambda(x)<\Lambda(\en)$. The latter two conditions are equivalent to $\en(x)\,\M(x) = \Lambda(\en)\,\M(x)$ for all $x$, thus proving the claim.
\end{proof}

To elucidate the previous proposition, suppose that a state ensemble $\en$ consists of $n$ equally probable quantum states. In this case the largest eigenvalue of each $\en(x)$ is at most $1/n$ and hence $\Lambda(\en) \leq 1/n$. 
Therefore, \eqref{eq:Pbound} gives $\Pg(\en) \leq d/n$. 
This bound has been called the \emph{basic decoding theorem} \cite{QPSI10} and it connects the Hilbert space dimension of a quantum system to its information capacity.

\subsection{Partition and property guessing games}\label{sec:partition}

There are two classes of guessing games, namely, partition and property guessing games, that are concrete in their goals but general enough to cover many applications.
In a \emph{partition guessing game} the input set $X$ is partitioned in some way and $Y$ labels the partitions of $X$.
For instance, we can take $X=\{1,\ldots,n\}$ and $Y=\{\text{even}, \text{odd}\}$.
The aim is to guess the correct quality of the input label, which is obviously less demanding than to guess the input label itself. 
Generally, suppose that $Y$ is an arbitrary set, $\upsilon:X \to Y$ is a function and let $X_y = \upsilon^{-1}(y)$ for all $y$.
Then, $(X_y)_{y\in Y}$ is a partition of $X$, i.e., a collection of subsets that are disjoint and whose union is $X$.
The associated score function $f_\upsilon$ is defined as
\begin{equation}\label{eq:partition}
f_\upsilon(x,y) = \delta_{\upsilon(x),y} = \begin{cases}
1 & \text{if $x\in X_y$} \\
0 &  \text{otherwise}
\end{cases}\,.
\end{equation}
This game has been studied in \cite{ZhYi02}, where it was called set discrimination of quantum states.
Another related score function $f_{\neg \upsilon}$ is defined as  $f_{\neg \upsilon}(x,y)=1-f_\upsilon(x,y)$.
In the special case when $X=Y$ and $\upsilon$ is the identity function, the score function $f_\upsilon$ is the standard discrimination score function $f_\delta$ and $f_{\neg \upsilon}$ is the standard antidiscrimination score function $1-f_\delta$.

Let $\upsilon:X \to Y$ be a function that determines a partition guessing game in the previously specified way.
The reduction formulas \eqref{eq:Delta} and \eqref{eq:enf_0} give $\Delta(\en,f_\upsilon)=1$ and 
\begin{equation}
\en_{f_\upsilon}(y) = \sum_{x \in X_y} \en(x) \, .
\end{equation}
We thus conclude that a partition guessing game can be recast as the usual discrimination game where the states are mixtures of the states in the blocks of the partition.

Partition guessing games are a special class of \emph{property guessing games}.  
While a partition divides a set $X$ into disjoint subsets, properties can have overlaps.
For instance, we can take $X=\{1,\ldots,n\}$, $Y=\{\text{small}, \text{large}\}$ and agree that `small' are numbers $x$ satisfying $x\leq \ceil{\tfrac{n+1}{2}}$  and  `large' are numbers $x$ satisfying $x\geq\floor{\tfrac{n+1}{2}}$.
In this case, the numbers $x$ with $\floor{\tfrac{n+1}{2}} \leq x\leq \ceil{\tfrac{n+1}{2}}$ have both properties.
Generally, suppose that $X$, $Y$ are arbitrary sets and $R \subset X \times Y$ is a relation.
The associated score function $f_R$ is defined as the indicator function of the set $R$, i.e.,
\begin{equation}
f_R(x,y) = 1_R(x,y) = \begin{cases}
1 & \text{if $xRy$} \\
0 &  \text{otherwise}
\end{cases}\,.
\end{equation}
Another related score function $f_{\neg R}$ is defined as  $f_{\neg R}(x,y)=1-f_R(x,y)$.
In the special case when $Y=\{S\subset X : \mo{S} = k\}$ and $R$ is the `belongs to' relation, the property guessing games defined via $f_R$ and $f_{\neg R}$ are the generalized (anti)discrimination games introduced at the end of Section \ref{sec:game}.

\section{Guessing games with posterior information}\label{sec:posterior}

\subsection{General scenario}\label{sec:scenario}

We will now expand the guessing game setup to cover later sent classical information.
Related formulations have been investigated earlier in \cite{BaWeWi08,GoWe10} and their differences to the current approach has been explained in \cite{CaHeTo18}, {where the following scheme was introduced in a more specialized form}.
In \emph{guessing games with posterior information}, the standard communication scenario is modified by adding one step to it. 
The starting point, known both to Alice and Bob, consists of finite sets $X$, $Y$, a score function $f:X\times Y \to [0,1]$, a finite set $T$ describing partial information, and conditional probabilities $\alpha(t\mid x)$ for all $t\in T$ and $x\in X$ relating partial information to input labels.
We can take $T=\{1,\ldots,m \}$ whenever it is convenient to label the elements of $T$ by integers, although this is not always the case as $T$ may not have a natural ordering (see Section \ref{sec:non-deterministic}).

\begin{figure}[h!]
\centering
\includegraphics[scale=0.7]{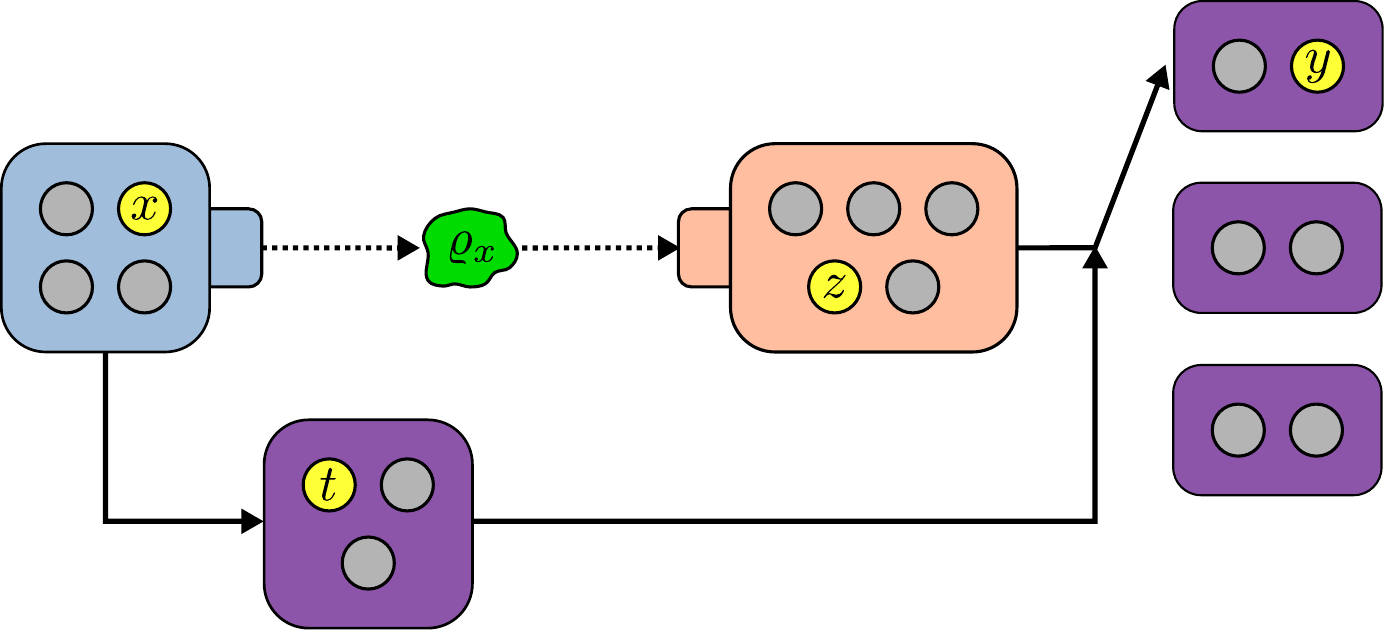}
\caption{In a guessing game with posterior information Bob receives Alice's partial information only after he has performed a measurement in the quantum state transmitted by her. He then postprocesses the obtained outcome trying to maximize the score of the game. \label{fig:post}
}
\end{figure}

The scenario has the following steps (see Figure \ref{fig:post}):
\begin{enumerate}[(i)]
\item Alice uses a state ensemble $\en$ with the label set $X$.
This means that she picks a label $x$ with probability $p(x)=\tr{\en(x)}$ and transmits the respective state $\varrho_x=\en(x)/\tr{\en(x)}$ to Bob.
\item Bob receives $\varrho_x$ and performs a measurement $\M$ with the outcome set $Z$. Bob obtains the outcome $z\in Z$ with probability $\tr{\varrho_x\,\M(z)}$.
\item Bob receives a classical message $t\in T$. 
This message depends on the input label $x$; Bob receives $t$ with probability $\alpha(t\mid x)$. This additional information can be sent by Alice, but it can have also another origin. 
The essential point is that Bob receives it after he has performed the measurement.
We call $\alpha$ the \emph{partial information map}.
\item Bob uses the additional information to post-process the obtained measurement outcome $z$ to an element $y\in Y$. 
For each $t\in T$, Bob can use a different post-processing matrix $\nu_t$ that relabels the outcome $z$ into $y$ with probability $\nu_t(y\mid z)$.
We denote $\nu: t \mapsto \nu_t$ and call this the \emph{post-processing map}.
The aim of Bob is to choose $y$ such that $f(x,y)$ is maximal. 
\end{enumerate}

Summarizing, a guessing game with posterior information is defined by a score function $f$ (the goal of the game) and a partial information map $\alpha$ (the additional aid for reaching the goal), while Alice's preparations are determined by $\en$ and Bob's guessing strategy is determined by a measurement $\M$ and post-processing map $\nu$.
The average score in the previously described scenario is
\begin{equation}\label{eq:Pgfpost}
\Efpost(\en;\M,\nu) =  \sum_{x,y,t,z} f(x,y)\, \alpha(t \mid x)\, \nu_t(y \mid z)\, \tr{\en(x)\,\M(z)}  
\end{equation}
and its maximal value is
\begin{equation}\label{eq:Ppost_f,alpha}
\Efpost(\en) =  \max_{\M,\nu}\,  \Efpost(\en;\M,\nu) \, ,
\end{equation}
where the optimization is over all measurements $\M$ and post-processing maps $\nu$. 
We remark that in \eqref{eq:Ppost_f,alpha} also the outcome set of $\M$ is allowed to vary. 
In particular, the fact that the maximum in \eqref{eq:Ppost_f,alpha} is attained is not immediate. 
However, we will prove in Section \ref{sec:standard} that this is indeed the case (see Proposition \ref{prop:max_barM}).

\subsection{Extreme cases}\label{sec:extreme}

There are two extreme cases of posterior information, those of telling everything or telling nothing.
It is illustrative to see how the scenario in these cases is simplified.

Firstly, Alice can tell the sent label $x$ to Bob as it is, in which case $T=X$ and $\alpha(t \mid x)=\delta_{t,x}$.
This means that the quantum prepare-and-measure part as well as the postprocessing  are obsolete and Bob -- as he learns $x$ -- can just choose $y_x$ such that $f(x,y_x)$ is maximal. 
Indeed, in this setting we have
\begin{align*}
\Efpost(\en;\M,\nu) & =  \sum_{x,y,z} f(x,y)\, \nu_x(y \mid z)\, \tr{\en(x)\,\M(z)}  \\
& \leq \sum_x f(x,y_x)\sum_{z}\bigg( \sum_y \nu_x(y \mid z)\bigg)\, \tr{\en(x)\,\M(z)}\\
& =\sum_x f(x,y_x)\,p(x)
\end{align*}
and the bound is achieved by choosing $\nu_x(y\mid z)=\delta_{y,y_x}$
The maximal average score is thus given as $\sum_x f(x,y_x)\,p(x)$.

Secondly, Alice can tell a posterior message $t$ that is independent of the original label, i.e., $\alpha(t \mid x) = \alpha(t)$. 
This kind of posterior information cannot help Bob.
In fact, from \eqref{eq:Pgfpost} we get
\begin{equation*}
\Efpost(\en;\M,\nu) = \Ef(\en;\M') \, ,
\end{equation*}
where
\begin{equation*}
\M'(y) = \sum_{z} \left( \sum_{t} \alpha(t)\,\nu_t(y\mid z) \right) \M(z) \, .
\end{equation*}
The post-processing that Bob might choose to perform can hence be included in the measurement and the guessing game reduces to that without posterior information, as expected.

Still a related special case is the one in which Alice may send useful posterior information but Bob is not taking advantage of it, i.e., Bob is post-processing his measurement outcome in a fixed manner. 
Formally, this means that the post-processing map $\nu:t\mapsto\nu_t$ is constant, hence the measurement
$$
\M''(y) = \sum_{z} \nu_t(y\mid z)\,\M(z)
$$
does not depend on $t$, and \eqref{eq:Pgfpost} takes the form
\begin{align*}
\Efpost(\en;\M,\nu) & =\sum_{x,y,t} f(x,y)\,\alpha(t \mid x)\, \tr{\en(x)\,\M''(y)} = \Ef(\en;\M'')\,.
\end{align*}
Choosing $Y=Z$ and $\nu_t(y\mid z) = \delta_{y,z}$ one has $\M''=\M$ and this confirms the intuitively clear fact that
\begin{equation}
\Efpost(\en) \geq \Ef(\en) 
\end{equation}
for any choice of $\alpha$, as Bob can always decide to ignore the posterior information.

\subsection{Deterministic posterior information}\label{sec:deterministic}\label{sec:non-overlapping}

As defined earlier, a partial information map $\alpha$ specifies how partial information relates to input labels.
Suppose that $\alpha(t\mid x)\in\{0,1\}$ for all $x,t$. 
Since $\sum_t \alpha(t \mid x) =1$, this means that for each $x\in X$ there is a unique $\tau(x)\in T$ such that $\alpha(\tau(x)\mid x) = 1$. 
Therefore, the input label $x$ specifies the later sent posterior information deterministically. 
By denoting $X_t = \tau^{-1}(t)$, the sets $(X_t)_{t\in T}$ constitute a partition of $X$ and $\alpha = \alpha_\tau$, where
\begin{equation}\label{eq:non-overlapping}
\alpha_\tau(t \mid x)= \delta_{\tau(x),t} = \begin{cases}
1 & \text{if $x\in X_t$} \\
0 &  \text{otherwise}
\end{cases}\,.
\end{equation}
We refer to this case as the case of \emph{deterministic posterior information}. For the task of state discrimination, this scenario has been discussed in \cite{CaHeTo18}.

As a paradigmatic exemplary case of the previously explained deterministic posterior information, we recall the discrimination task presented in \cite{AkKaMa19}, where $|X|=|Y|=4$ and $|T|=2$.
In this guessing game the set $X$ can be chosen to contain four symbols $\{\clubsuit, \spadesuit, \diamondsuit, \heartsuit\}$, and Alice chooses the input label among them with uniform probability. 
She uses a qutrit system to send her message to Bob, and the respective (pure) qutrit states correspond to the unit vectors
\begin{equation*}
\varrho_{\clubsuit} \sim \frac{1}{\sqrt{2}} \left(\begin{array}{c}1 \\ 1 \\0\end{array}\right) , \ \varrho_{\spadesuit} \sim \frac{1}{\sqrt{2}} \left(\begin{array}{c}1 \\ -1 \\0\end{array}\right) , \ \varrho_{\diamondsuit}\sim \frac{1}{\sqrt{2}} \left(\begin{array}{c}1 \\ 0 \\ 1\end{array}\right) , \ \varrho_{\heartsuit}\sim \frac{1}{\sqrt{2}} \left(\begin{array}{c}1 \\ 0 \\ -1\end{array}\right) .
\end{equation*}
These are four states of a three dimensional system, hence there is no measurement that would perfectly discriminate them. 
In fact, Proposition \ref{prop:Pbound} and the discussion after it implies that $\Pg(\en) \leq 3/4$ for any uniformly distributed four qutrit states.

However, we are considering discrimination with posterior information and Bob knows that after he has performed the measurement, Alice will inform him about the color of the symbol (black for $\{\clubsuit, \spadesuit\}$ and red for $\{\diamondsuit, \heartsuit\}$).
In our notation, this means that the partition of $X$ is $X_{\textrm{black}}=\{\clubsuit, \spadesuit\}$ and $X_{\textrm{red}}=\{\diamondsuit, \heartsuit\}$.
The measurement $\M$ that Bob wisely decides to use is 
\begin{align*}
\M(1) & =\frac{1}{4}\left(\begin{array}{ccc}1 & 1 & 1 \\1 & 1 & 1 \\1 & 1 & 1\end{array}\right) , & \M(2) & =\frac{1}{4}\left(\begin{array}{ccc}1 & 1 & -1 \\1 & 1 & -1 \\-1 & -1 & 1\end{array}\right) , \\
\M(3) & =\frac{1}{4}\left(\begin{array}{ccc}1 & -1 & 1 \\-1 & 1 & -1 \\1 & -1 & 1\end{array}\right) , & \M(4) & =\frac{1}{4}\left(\begin{array}{ccc}1 & -1 & -1 \\-1 & 1 & -1 \\-1 & -1 & 1\end{array}\right) .
\end{align*}
This leads to the probability distributions
\begin{align*}
\tr{\varrho_{\clubsuit}\,\M(\cdot)} & = \left(\half,\half,0,0\right) \, , \\
\tr{\varrho_{\spadesuit}\,\M(\cdot)} & = \left(0,0,\half,\half\right) \, , \\
\tr{\varrho_{\diamondsuit}\,\M(\cdot)} & = \left(\half,0,\half,0\right) \, , \\
\tr{\varrho_{\heartsuit}\,\M(\cdot)} & = \left(0,\half,0,\half\right) \, .
\end{align*}
From these probabilities we confirm that Bob can indeed infer the correct input label if he gets the color of the input symbol as a posterior information. 
For example, if the outcome is $z=2$, then Bob needs to 
post-process it to $\clubsuit$ if the color is black, and to $\heartsuit$ if the color is red.

\subsection{Non-deterministic posterior information}\label{sec:non-deterministic}

We say that posterior information is {\em non-deterministic} if $0<\alpha(t\mid x)<1$ at least for some $x$ and $t$. 
This means that for some $x$ there are at least two possible labels $t$ and $t'$ that can occur as partial information when $x$ is the sent input label{, and thus}
Alice makes a random choice between some alternatives.

A paradigmatic exemplary case of non-deterministic posterior information is the exclusion of wrong options.
Let us set $T=X$ and define
\begin{equation}\label{eq:ruling-out}
\alpha_{\rm ex}(t\mid x) = (|X|-1)^{-1} (1-\delta_{x,t}) \,. 
\end{equation}
This partial information map means that Alice announces one wrong option $t$ after Bob has performed his measurement, and she picks it with uniform probability within the set $X\setminus\{x\}$. More generally, we can fix any positive integer $k < |X|$ and define
\begin{equation}
T = \{S\subset X : \mo{S} = k\}\,,\qquad\quad \alpha_{\rm ex}(S\mid x) = |X|\,[\,|T|\,(|X|-k)\,]^{-1}\,
1_{X\setminus S}(x) \,,
\end{equation}
where $|T| = |X|!\,[\,k!\,(|X|-k)!\,]^{-1}$ and the normalization constant of $\alpha_{\rm ex}$ is the inverse cardinality of the set $T_x = \{S\in T : x\notin S\}$. This choice of $\alpha$ means that Alice announces a collection of $k$ wrong options $S=\{x_1,\ldots,x_k\}$, and she picks it with uniform probability within the set $T_x$.

\section{{Reduction} to usual state discrimination games}\label{sec:reduction}

\subsection{Standard form of guessing games with posterior information}\label{sec:standard}

As we have previously seen,  Bob's guessing strategy is determined by a measurement $\M$ and post-processing map $\nu$.
There is a certain freedom in choosing $\M$ and $\nu$, still leading to the same average score for a given state ensemble $\en$.
To see this, we write the average score \eqref{eq:Pgfpost} as
\begin{equation}\label{eq:compatible}
\Efpost(\en;\M,\nu) 
=  \sum_{x,y,t} f(x,y)\, \alpha(t \mid x)\, \tr{\en(x)\,\N_t(y)} \, ,
\end{equation}
where $\N_t$ are the post-processed measurements defined as
\begin{equation}\label{eq:compatible-N}
\N_t(y) = \sum_{z} \nu_t(y \mid z) \, \M(z)\,.
\end{equation}
Thus, different measurements $\M$ and post-processing maps $\nu$ which yield the same measurements $\N_t$ in \eqref{eq:compatible-N} lead to equal average scores.

Given a collection of measurements $(\N_t)_{t \in T}$, all with the same outcome set $Y$, we recall that the collection is called {\em compatible} if each $\N_t$ can be written as in \eqref{eq:compatible-N} for some choice of $\M$ and $\nu$ \cite{HeMiZi16}. 
Otherwise, one says that the collection is {\em incompatible}. 
As a consequence of \eqref{eq:Ppost_f,alpha} and \eqref{eq:compatible}, we can write
\begin{equation*}
\Efpost(\en) = \max\bigg\{ \sum_{x,y,t} f(x,y)\, \alpha(t \mid x)\, \tr{\en(x)\,\N_t(y)} : (\N_t)_{t \in T} \text{ is compatible} \bigg\}\,.
\end{equation*}
The compatibility constraint guarantees that the two measurement scenarios -- using $\M$ and post-processing, or using the collection $(\N_t)_{t\in T}$ -- are equivalent. In fact, without the compatibility constraint, the scenario with many measurements becomes a guessing game with \emph{prior} information. We come back to this point in Section \ref{sec:prior}.

The outcome set of $\M$ in the definition of compatibility of $(\N_t)_{t \in T}$ is not fixed and it can be arbitrary.
However, every compatible collection of measurements has a {\em joint measurement}, i.e., a measurement defined on their product outcome set and giving them as marginals \cite{AlCaHeTo09}.
In the current context, this means that we can always switch from $\M$ to a measurement with the outcome set $Y^T$ and to a fixed post-processing map, defined as
\begin{equation}\label{eq:def_pi}
\pi_t(y\mid \phi)= \delta_{y,\phi(t)} =
\begin{cases}
1 & \text{ if $y=\phi(t)$} \\
0 & \text{ if $y\neq\phi(t)$}
\end{cases}\,.
\end{equation}
(Here and in the following we use the customary notation $Y^T$ for the set of all maps $\phi:T\to Y$. If $T=\{1,\ldots,m\}$, then $Y^T$ is identified with the product set $Y^m$ canonically. The functional notation is convenient especially when $T$ does not have any natural ordering.)
In fact, starting from $\M$ and $\nu$, we define a measurement $\bar{\M}_{\nu}$ with the outcome set $Y^T$ as
\begin{equation}
\bar{\M}_{\nu}(\phi) = \sum_z \M(z) \prod_t \nu_t(\phi(t) \mid z)
\end{equation}
and then we have
\begin{equation*}
\begin{aligned}
\sum_\phi \pi_t (y \mid \phi)\,\bar{\M}_{\nu}(\phi) & = \sum_z \M(z) \,\nu_t(y \mid z) \prod_{t'\neq t} \sum_{y'} \nu_{t'}(y' \mid z) \\
& = \sum_{z} \nu_t(y \mid z)\,\M(z)\,,
\end{aligned}
\end{equation*}
which means that the post-processed measurements \eqref{eq:compatible-N} are the \emph{marginals} of $\bar{\M}_{\nu}$. In particular, 
\begin{equation}\label{eq:Epost(M,nu)=Epost(barM,pi)}
\Efpost(\en;\M,\nu)=\Efpost(\en;\bar{\M}_{\nu},\pi) \, .
\end{equation}
The importance of this transition from $\M$ and $\nu$ to $\bar{\M}_{\nu}$ and $\pi$ is that for the latter pair the outcome set is fixed and so is also the post-processing map. We thus reach the following conclusion.

\begin{proposition}\label{prop:max_barM}
The maximum in \eqref{eq:Ppost_f,alpha} is attained, and
\begin{equation}\label{eq:max_barM}
\Efpost(\en) = \max_{\bar{\M}}\, \Efpost(\en;\bar{\M},\pi) \,,
\end{equation}
where the optimization is over all measurements $\bar{\M}$ with the outcome set $Y^T$.
\end{proposition}
\begin{proof}
Clearly, $\Efpost(\en;\bar{\M},\pi)\leq\Efpost(\en)$ for all $\bar{\M}$, and \eqref{eq:max_barM} is then a consequence of the bound
$$
\Efpost(\en;\M,\nu) \leq \max_{\bar{\M}}\, \Efpost(\en;\bar{\M},\pi)
$$
following from \eqref{eq:Epost(M,nu)=Epost(barM,pi)}. The maxima in \eqref{eq:Ppost_f,alpha} and \eqref{eq:max_barM} are attained, since the measurements with the outcome set $Y^T$ form a compact set and in \eqref{eq:max_barM} the post-processing map $\pi$ is fixed.
\end{proof}

As a result, if the objective is to optimize the average score of a guessing game with posterior information that has $Y$ as the output label set and $T$ as the partial information set, it is enough to consider guessing strategies of the following \emph{standard form}:
\begin{itemize}
\item Bob is using a measurement $\bar{\M}$ with the outcome set $Y^T$. From the obtained measurement outcome $\phi$, he chooses $\phi(t)$ based on the posterior information $t \in T$.
\end{itemize}
This general formulation is useful for proving results in the subsequent sections.

\subsection{Reduction theorem}

In the following we present the basic steps how the maximal average score in a guessing game with posterior information can be calculated. 
Our approach is related but more general than a result presented in \cite{CaHeTo18}.
The main point is that a standard form guessing game with posterior information can be translated to a usual state discrimination task.
We first recall from Section \ref{sec:standard} than in the standard form Bob's measurement is defined on the product outcome set $Y^T$.
For any measurement $\bar{\M}$ with the product outcome set $Y^T$ and for the post-processing map $\pi$ defined in \eqref{eq:def_pi}, the average score \eqref{eq:Pgfpost} can be rewritten as
\begin{equation}\label{eq:aux_Pgfpost}
\begin{aligned}
\Efpost(\en;\bar{\M},\pi) & = \sum_{x,y,t,\phi} f(x,y)\, \alpha(t \mid x)\,\pi_t(y\mid\phi)\, \tr{\en(x)\,\bar{\M}(\phi)} \\
& = \sum_\phi {\rm tr}\bigg[\bigg( \sum_{x,y,t} f(x,y) \, \alpha(t \mid x)\, \delta_{y,\phi(t)}\, \en(x) \bigg)\,\bar{\M}(\phi)\bigg] \\
& = \sum_\phi {\rm tr}\bigg[\bigg( \sum_{x,t} f(x,\phi(t)) \, \alpha(t \mid x)\, \en(x) \bigg)\,\bar{\M}(\phi)\bigg] \\
& = \mo{Y}^{\mo{T}-1}\Delta(\en,f)\ \Pg\left(\en_{f,\alpha}\, ;\, \bar{\M}\right) \,.
\end{aligned}
\end{equation}
In the last expression, $\Delta(\en,f)$ is the constant defined in \eqref{eq:Delta}, while $\en_{f,\alpha}$ is a new state ensemble with the label set $Y^T$, which extends the auxiliary state ensemble \eqref{eq:enf_0} to the scenario with posterior information.
Under the presumption $\Delta(\en,f)\neq 0$ it is defined as
\begin{equation}\label{eq:enf-alpha}
\en_{f,\alpha} (\phi) = \big(\mo{Y}^{\mo{T}-1}\Delta(\en,f)\big)^{-1} \sum_{x,t} f(x,\phi(t)) \, \alpha(t \mid x) \, \en(x) \, .
\end{equation}
(In the case $\Delta(\en,f) = 0$ we can set, for instance, $\en_{f,\alpha} (\phi) = \big(d\,|Y|^{|T|}\big)^{-1}\,\id$ and then the following formulae cover also this situation.)
The normalization constant before the sum in \eqref{eq:enf-alpha} is due to the fact that
\begin{equation*}
\begin{aligned}
\sum_\phi{\rm tr}\bigg[\sum_{x,t} f(x,\phi(t)) \, \alpha(t \mid x)\, \en(x)\bigg] & = \mo{Y}^{\mo{T}-1} \sum_{x,y,t} f(x,y) \, \alpha(t \mid x) \, p(x) \\
& = \mo{Y}^{\mo{T}-1} \Delta(\en,f) \,.
\end{aligned}
\end{equation*}
The main purpose of introducing the auxiliary state ensemble is summarized in the following statement.

\begin{theorem}\label{thm:reduction}
For any $\en,\alpha$ and $f$, we have 
\begin{equation}\label{eq:reduction}
\Efpost(\en) = \mo{Y}^{\mo{T}-1}\Delta(\en,f)\ \Pg(\en_{f,\alpha}) \, .
\end{equation}
\end{theorem}
\begin{proof}
The claim follows by combining \eqref{eq:max_barM} and \eqref{eq:aux_Pgfpost}.
\end{proof}

We remark that the definition of the auxiliary state ensemble $\en_{f,\alpha}$ is consistent with the earlier definition of the auxiliary state ensemble \eqref{eq:enf_0}. Indeed, if the posterior information is trivial, then $|T|=1$, implying that $\en_{f,\alpha} = \en_f$ and \eqref{eq:aux_Pgfpost}, \eqref{eq:reduction} reduce to \eqref{eq:aux_Pgfpost_1}, \eqref{eq:aux_Pgfpost_2}, respectively.

\subsection{State discrimination with deterministic posterior information}\label{ex:non-overlapping-reduction}

State discrimination with deterministic posterior information is a prototypical example of the discussed scenario and in Section \ref{sec:deterministic} we recalled one concrete case of that type. 
For this class of games the auxiliary state ensemble has a simple form.
To see it, let $T = \{1,\ldots,m\}$, fix a function $\tau:X\to T$, set $X_t=\tau^{-1}(t)$ and define the partial information map $\alpha_\tau$ as in \eqref{eq:non-overlapping}. Moreover, let $Y=X$ and fix the standard discrimination score function $f=f_\delta$.
The auxiliary state ensemble \eqref{eq:enf-alpha} becomes
\begin{equation}\label{eq:enfalpha_discr2}
\en_{f_\delta,\alpha_\tau} (x_1,\ldots,x_m) = |X|^{1-m} \sum_{\substack{t\text{ s.t.}\\ x_t\in X_t}} \en(x_t) \,,
\end{equation}
where we write elements $\phi\in X^T$ as ordered $m$-tuples $(x_1,\ldots,x_m)$ with $x_t = \phi(t)$. 

We remark that this case was already studied in \cite{CaHeTo18}, where it was proved that $\Epost_{f_\delta,\alpha_\tau}(\en) = \Delta'\ \Pg(\enf)$ for another definition of the constant $\Delta'$ and the auxiliary state ensemble $\enf$ (see equations (22) and (23) therein). The difference between the ensembles $\en_{f_\delta,\alpha_\tau}$ and $\enf$ is in the respective label sets, which are the product {set} $X^m$ for the former ensemble and $X_1\times\ldots\times X_m$ for the latter one.
Actually, up to the constant factor $\Delta'\,|X|^{1-m}$, the state ensemble $\enf$ coincides with the restriction of $\en_{f_\delta,\alpha_\tau}$ to the set $X_1\times\ldots\times X_m$. 
Therefore, we see that in this particular case there is a certain amount of redundancy in employing the auxiliary state ensemble $\en_{f,\alpha}$ to evaluate $\Efpost(\en)$.

\subsection{State discrimination with the exclusion of {wrong options}}

State discrimination with the random exclusion of one wrong option  was discussed in Section \ref{sec:non-deterministic}. 
Suppose that $T=X$ and $\alpha=\alpha_{\rm ex}$ is the partial information map \eqref{eq:ruling-out}. 
The auxiliary state ensemble \eqref{eq:enf-alpha} becomes
\begin{equation}\label{eq:ruling-out_enf}
\begin{aligned}
\en_{f,\alpha_{\rm ex}}(\phi) & = C\,\sum_t \sum_x f(x,\phi(t))\,(1-\delta_{x,t})\,\en(x) \\
& = C\,\sum_{y\in\phi(X)} \sum_{\substack{t\text{ s.t.}\\ \phi(t) = y}}\, \sum_x f(x,y)\,(1-\delta_{x,t})\, \en(x) \\
& = C\,\sum_{x,y} f(x,y)\,\mo{\phi^{-1}(y)\setminus\{x\}}\,\en(x) \,,
\end{aligned}
\end{equation}
where $1/C=(\mo{X}-1)\,|Y|^{|T|-1}\Delta(\en,f)$.
We observe that the dependence on the outcome $\phi$ is only in the cardinalities $\mo{\phi^{-1}(y)\setminus\{x\}}$ appearing in the last line of \eqref{eq:ruling-out_enf}. These are integer numbers between $0$ and $|X|-1$ such that $\sum_x\mo{\phi^{-1}(y)\setminus\{x\}}\leq |X|$.

We can perform a similar computation for the case in which one excludes more than one {wrong} option. Using the same notation as in Section \ref{sec:non-deterministic}, we find that
\begin{equation}\label{eq:ruling-out_enf2}
\begin{aligned}
\en_{f,\alpha_{\rm ex}}(\phi) & = C\,\sum_{x,y} f(x,y)\,\mo{\phi^{-1}(y)\cap T_x}\,\en(x) \,,
\end{aligned}
\end{equation}
where {$1/C = |T|\,|X|^{-1}(|X|-k)\,|Y|^{|T|-1} \Delta(\en,f)$.}

\section{Guessing games with prior information}\label{sec:prior}

\subsection{General scenario}\label{sec:prior-gen}

\begin{figure}[h!]
\centering
\includegraphics[scale=0.7]{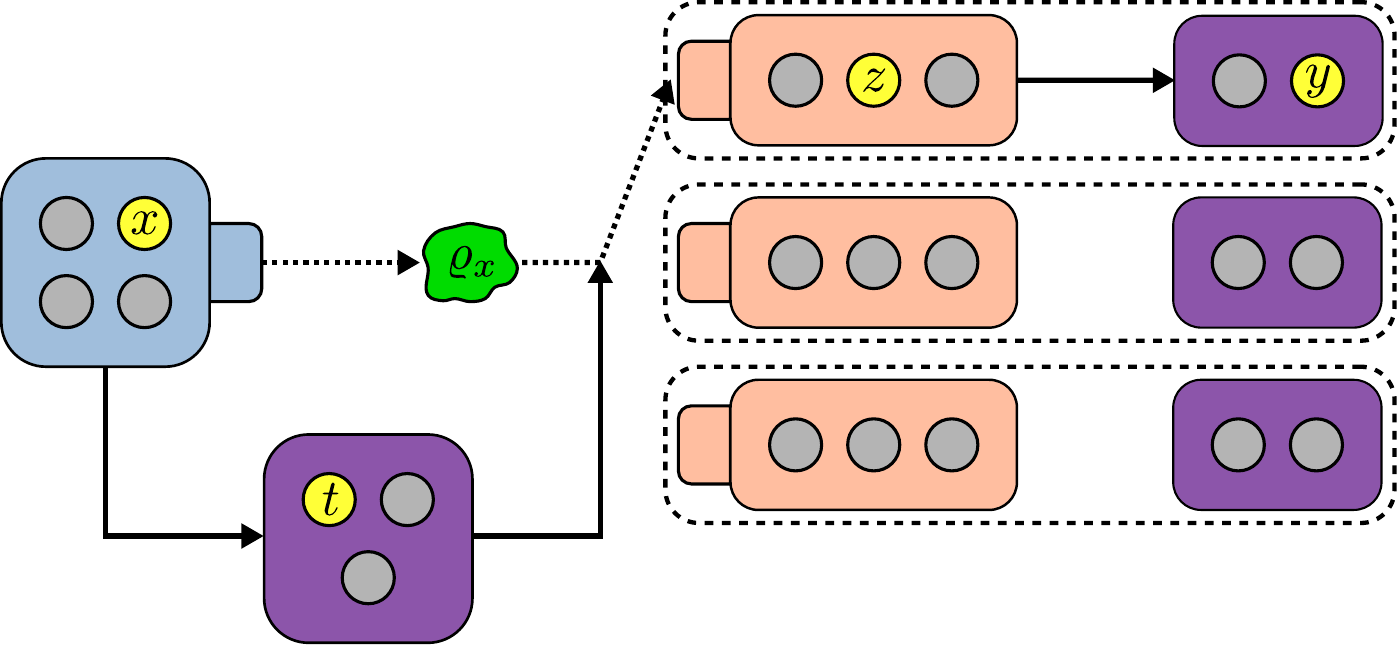}
\caption{In a guessing game with prior information Bob arranges his measurement after he receives Alice's partial information. The postprocessing of the obtained outcome can now be included in the measurement itself. In this scenario Bob is allowed to optimize his measurement in order to get the highest score. \label{fig:pre}}
\end{figure}

We recall that $\Efpost(\en)$ denotes the best achievable average score when the optimization is over all measurements $\M$ and post-processing maps $\nu$. 
In Section \ref{sec:standard} we saw that finding $\Efpost(\en)$ is equivalent to optimizing the sum in \eqref{eq:compatible} over the compatible collections of $|T|$ measurements with the outcome set $Y$.
One can obviously write such a sum also without the assumption of compatibility, but ignoring this constraint may lead to a larger maximal average score than $\Efpost(\en)$.
In fact, the usage of the additional information $t$ for the choice of the measurement $\N_t$ means that $t$ is used prior the measurement happens.
We call this different scenario a \emph{guessing game with prior information} (see Figure \ref{fig:pre}), and we write
\begin{equation}\label{eq:pre}
\Efpre(\en;( \N_t )_{t \in T})  =  \sum_{x,y,t} f(x,y)\, \alpha(t \mid x)\, \tr{\en(x)\,\N_t(y)}
\end{equation}
for its average score.

The maximal average score is
\begin{equation}\label{eq:sup_any}
\Efpre(\en) = \max\big\{ \Efpre(\en;( \N_t )_{t \in T}) : ( \N_t )_{t \in T} \text{ is any collection of measurements} \big\}\,.
\end{equation}
The evaluation of $\Efpre(\en)$ boils down to determining the maximal average scores of $|T|$ different guessing games of the usual type. 
To see this, we introduce the total probability
\begin{equation}
q(t) = \sum_x \alpha(t\mid x)\,p(x)
\end{equation}
and, whenever $q(t)$ is nonzero, we define the {\em conditional state ensemble} $\en_t$ as follows:
\begin{equation}\label{eq:ent}
\en_t(x) = q(t)^{-1}\alpha(t\mid x)\,\en(x)\,.
\end{equation}
With the above definition, we can rewrite \eqref{eq:pre} as
\begin{equation}
\Efpre(\en;(\N_t)_{t\in T}) = \sum_t q(t)\,\Ef(\en_t;\N_t)
\end{equation}
and by combining \eqref{eq:def_E(E)} and \eqref{eq:sup_any}, we then obtain
\begin{equation}\label{eq:convex}
\Efpre(\en) = \sum_t q(t)\,\Ef(\en_t) \, .
\end{equation}
Thus, the maximal average score with prior information $\Efpre(\en)$ is a convex sum of maximal average scores $\Ef(\en_t)$ for different $t$.
It can hence be evaluated by means of the techniques of usual minimum error state discrimination, applied to each conditional state ensemble $\en_t$. The definition of $\en_t$ is subject to the same remarks as those after the introduction of the auxiliary state ensemble in \eqref{eq:enf_0}. Note that in the present case the label sets of $\en_t$ and $\en$ coincide, and that their states are essentially the same. Indeed, $\varrho_{t,x} = \en_t(x)/\tr{\en_t(x)}$ and $\varrho_x = \en(x)/\tr{\en(x)}$ are equal for all $t$ and $x$ such that $q(t)\neq 0$ and $\en_t(x)\neq 0$. 
On the other hand, the probabilities $p_t(x) = \tr{\en_t(x)}$ and $p(x) = \tr{\en(x)}$ may be different in general.

\subsection{Detection of incompatibility}\label{sec:incomp}

Summarizing the earlier discussion, a score function $f$ and a partial information map $\alpha$ define two different guessing games as the partial information can be delivered to Bob either before or after he is performing a measurement.
If Bob can access this information before, then the average score is $\Efpre(\en;( \N_t )_{t \in T})$ given in \eqref{eq:pre}.
While if Bob gets the information only later and his usage of it is therefore limited to post-processing the measurement outcomes, then we are back in the guessing game with posterior information, and the maximal average score is
\begin{equation}\label{eq:sup_compatible}
\Efpost(\en) = \max\big\{ \Efpre(\en;( \N_t )_{t \in T}) : ( \N_t )_{t \in T} \text{ is compatible} \big\}
\end{equation}
as discussed in Section \ref{sec:standard}. 
We thus see that the difference of the two games is really about (in)compatibility of measurements.
The following result, first proved in \cite{CaHeTo18} for a more restricted scenario, is based on these observations.

\begin{proposition}\label{prop:greater->incomp}
If $\Efpre(\en;( \N_t )_{t \in T}) > \Efpost(\en)$, then $( \N_t )_{t \in T}$ is incompatible.
\end{proposition}

The opposite question is: if $( \N_t )_{t \in T}$ is a collection of incompatible measurements, how can we detect their incompatibility by performing a guessing game?
This means that we compare the average score $\Efpre(\en;( \N_t )_{t \in T})$ to the maximal average score with posterior information, $\Efpost(\en)$. 
The first one can even be calculated from experimental data if $\N_t$ are real devices, whereas $\Efpost(\en)$ can be determined or at least upper bounded analytically (more about that in later sections).
This question has been studied from various different angles in \cite{CaHeTo19, UoKrShYuGu19,SkSuCa19,BuChZh20,Kuramochi20} and important findings have been reported.
One statement is the following (see Theorem 2 in \cite{CaHeTo19}).

\begin{theorem}\label{prop:incomp->greater}
Let $X=Y\times T$ and $\upsilon:X\to Y$, $\tau:X\to T$ be the projections of $X$ onto the respective factors. Moreover, fix the score function $f_\upsilon$ and the partial information map $\alpha_\tau$ as in \eqref{eq:partition} and  \eqref{eq:non-overlapping}, respectively. 
Then, for any incompatible collection of measurements $( \N_t )_{t \in T}$ with the outcome set $Y$, there exists a state ensemble $\en$ with the label set $X$ such that $\Epre_{f_\upsilon,\alpha_\tau}(\en;( \N_t )_{t \in T}) > \Epost_{f_\upsilon,\alpha_\tau}(\en)$.
\end{theorem}

\begin{proof}
The proof is a straightforward adaptation of the argument provided in \cite{CaHeTo19}.\\
Let $\mathcal{V}$ be the linear space of all collections $(F_t)_{t\in T}$ of operator valued functions $F_t : Y\to\lh$. Any collection of measurements $(\N_t)_{t\in T}$ with the outcome set $Y$ is an element of $\mathcal{V}$, and all collections which are compatible constitute a compact convex subset $\mathcal{C}\subset\mathcal{V}$. Indeed, by the discussion in Section \ref{sec:standard}, a collection $(\N_t)_{t\in T}$ is compatible if and only if each measurement $\N_t$ is obtained as the marginal of a joint measurement, and joint measurements form a compact convex subset of the linear space of all $\lh$-valued functions on $Y^T$. Now, suppose $(\N_t)_{t\in T}$ is an incompatible collection of measurements. By a standard separation argument, there exists a hyperplane in $\mathcal{V}$ which separates $(\N_t)_{t\in T}$ from $\mathcal{C}$. Equivalently, one can find $(F_t)_{t\in T}\in\mathcal{V}$ and $\kappa\in\R$ such that, by defining
$$
\xi\big((\N'_t)_{t\in T}\big) = \kappa - \sum_{y,t} \tr{F_t(y)\,\N'_t(y)}
$$
for all collections of measurements $(\N'_t)_{t\in T}$, the inequality $\xi\geq 0$ holds on the set $\mathcal{C}$, while $\xi\big((\N_t)_{t\in T}\big) < 0$ for the incompatible collection $(\N_t)_{t\in T}$. Fix $\mu>0$ satisfying $F_t(y)+(\mu/2)\,\id \geq 0$ for all $y,t$, and let $1/\lambda = \sum_{y,t} {\rm tr}\big[F_t(y)+\mu\,\id\big]>0$. Define
$$
\en(y,t) = \lambda\,\big(F_t(y)+\mu\,\id\big) \,.
$$
It is easy to check that $\en$ is a state ensemble with the label set $X$. Moreover,
$$
\Epre_{f_\upsilon,\alpha_\tau}(\en;( \N'_t )_{t \in T}) = -\lambda\,\xi\big((\N'_t)_{t\in T}\big) + \kappa'\,,
$$
where $\kappa' = \lambda\,(\kappa + d\,\mu\,|T|)$. By \eqref{eq:sup_compatible}, it follows that
\begin{align*}
\Epost_{f_\upsilon,\alpha_\tau}(\en) & = -\lambda\,\min\big\{ \xi\big((\N'_t)_{t\in T}\big) : ( \N'_t )_{t \in T} \in\mathcal{C} \big\} + \kappa' \\
& < -\lambda\,\xi\big((\N_t)_{t\in T}\big) + \kappa' = \Epre_{f_\upsilon,\alpha_\tau}(\en;( \N_t )_{t \in T})
\end{align*}
as claimed in the theorem.
\end{proof}

We underline that, in order to detect all incompatible collections of measurements with the outcome set $Y$ in a guessing game with partial information from the set $T$, Theorem \ref{prop:incomp->greater} requires a sufficiently large label set $X$, namely, $|X| = |Y|\,|T|$.

Combined together, Proposition \ref{prop:greater->incomp} and Theorem \ref{prop:incomp->greater} lead to the conclusion that a collection $( \N_t )_{t \in T}$ is incompatible if and only if there is a guessing game such that $\Efpost(\en;( \N_t )_{t \in T}) > \Efpost(\en)$ for some choice of $f$, $\alpha$ and $\en$.
It appears that the full realm of guessing games with posterior information has not yet been investigated from the viewpoint of incompatibility detection and there are several open questions.
For instance, when a given class of such guessing games is enough to detect all incompatible collections of measurements?
In particular, is it possible to use smaller state ensembles and still be able to detect incompatibility?
Further, what is the condition for a pair of a score function $f$ and a partial information map $\alpha$ to detect some incompatible pair?
We leave these questions for future investigations.

\subsection{Quantum versus classical information}\label{sec:classical}

Proposition \ref{prop:greater->incomp} and Theorem \ref{prop:incomp->greater} also point out a fundamental difference between quantum and classical theory: while quantum theory admits guessing games in which prior information gives an advantage over posterior information, in classical theory the two scenarios are equivalent.
In terms of the maximal average scores \eqref{eq:sup_any} and \eqref{eq:sup_compatible}, this amounts to say that for any classical state ensemble $\en$, we have $\Efpre(\en) = \Efpost(\en)$ for all $f$ and $\alpha$. To give a precise explanation of this statement, we recall that the states of a (finite dimensional) classical system are just probability distributions on a fixed finite set $H$. Denoting by $\ell(\cdot)$ the linear space of all complex functions on a given set, measurements on $H$ with the outcome set $Z$ are described by linear positive maps $\M^{\scriptscriptstyle\wedge} {:\ell(H)}\to\ell(Z)$ which send the probability distributions on $H$ into those on $Z$. The general structure is
\begin{equation}
\big[\M^{\scriptscriptstyle\wedge}(g)\big](z) = \sum_h \mu(z\mid h)\,g(h) \quad \forall g\in\ell(H) \,,
\end{equation}
where $\mu(z\mid h)$ are conditional probabilities uniquely determined by the measurement $\M^{\scriptscriptstyle\wedge}$. For classical guessing games, everything goes as in the quantum case up to replacing the Born rule $\tr{\en(x)\,\M(z)}$ with the probabilities $\big[\M^{\scriptscriptstyle\wedge}(\en(x))\big](z)$ inside the expressions of the average scores. In classical theory, any collection $(\N^{\scriptscriptstyle\wedge}_t)_{t\in T}$ of measurements with the outcome set $Y$ is compatible. 
Indeed, if
\begin{equation}
\big[\N^{\scriptscriptstyle\wedge}_t(g)\big](y) = \sum_h \nu_t(y\mid h)\,g(h) \,,
\end{equation}
then each $\N^{\scriptscriptstyle\wedge}_t$ is the marginal of the following measurement $\bar{\M}^{\scriptscriptstyle\wedge}_\nu$ with the product outcome set $Y^T$
\begin{equation}
\big[\bar{\M}^{\scriptscriptstyle\wedge}_\nu (g)\big](\phi) = \sum_h g(h) \prod_t \nu_t(\phi(t) \mid h)\,.
\end{equation}
In particular, for all $f$, $\alpha$ and $\en$, we have $\Efpre(\en;(\N^{\scriptscriptstyle\wedge}_t)_{t \in T}) = \Efpost(\en;\bar{\M}^{\scriptscriptstyle\wedge}_\nu,\pi)$, where $\pi$ is the post-processing map defined in \eqref{eq:def_pi}. This implies that the probability $\Efpre(\en;(\N^{\scriptscriptstyle\wedge}_t)_{t \in T})$ can not exceed the bound $\Efpost(\en)$, as claimed.

When the equality $\Efpre(\en) = \Efpost(\en)$ holds, we say that \emph{the timing of partial information is irrelevant} for the state ensemble $\en$ in the guessing game with score function $f$ and partial information map $\alpha$.
As we have just seen, this is always the case for guessing games based on classical systems. 
It is still true for quantum state ensembles which are diagonal with respect to a fixed reference basis of the system Hilbert space, as shown in the following statement.

\begin{theorem}\label{thm:classical}
Suppose $\en$ is a state ensemble such that the operators $\en(x)$ and $\en(x')$ commute for all $x$, $x'$ belonging to the label set of $\en$. Then, the timing of partial information is irrelevant for $\en$ in all guessing games.
\end{theorem}

\begin{proof}
We show that for all collections of measurements $(\N_t)_{t\in T}$ there exists a compatible collection $(\N'_t)_{t\in T}$ such that
\begin{equation}\label{eq:equ_proof}\tag{$\ast$}
\Efpre(\en;(\N_t)_{t\in T}) = \Efpre(\en;(\N'_t)_{t\in T})\,,
\end{equation}
and then the claim follows from \eqref{eq:sup_any} and \eqref{eq:sup_compatible}. Let $(\varphi_h)_{h\in H}$ be an orthonormal basis of $\hh$ which diagonalizes all the operators $\en(x)$, $x\in X$. We define two linear maps $\Phi_{\rm meas}:\lh\to\ell(H)$ and $\Phi_{\rm prep}:\ell(H)\to\lh$ as follows:
$$
\big[\Phi_{\rm meas}(\varrho)\big](h) = {\rm tr}\big[\kb{\varphi_h}{\varphi_h}\,\varrho\big]\,,\qquad\qquad \Phi_{\rm prep}(g) = \sum_h g(h)\,\kb{\varphi_h}{\varphi_h}\,.
$$
The state ensemble $\en$ is invariant with respect to the composed map $\Phi_{\rm prep}\circ\Phi_{\rm meas}$, that is, $\Phi_{\rm prep}\big(\Phi_{\rm meas}(\en(x))\big) = \en(x)$ for all $x$. Let $\N^{\scriptscriptstyle\wedge}_t$ be the classical measurement on $H$ with the outcome set $Y$ which is given by
$$
\big[\N^{\scriptscriptstyle\wedge}_t (g)\big](y) = \tr{\Phi_{\rm prep}(g)\,\N_t(y)} \,.
$$
We have
\begin{align*}
\tr{\en(x)\,\N_t(y)} & = \tr{\Phi_{\rm prep}\big(\Phi_{\rm meas}(\en(x))\big)\,\N_t(y)} = \big[\N^{\scriptscriptstyle\wedge}_t \big(\Phi_{\rm meas}(\en(x))\big)\big](y) \\
& = \dual{(\N^{\scriptscriptstyle\wedge}_t\circ\Phi_{\rm meas})(\en(x))}{\delta_y} = \tr{\en(x)\,(\N^{\scriptscriptstyle\wedge}_t\circ\Phi_{\rm meas})^*(\delta_y)} \,,
\end{align*}
where $\dual{g}{\gamma}=\sum_y g(y)\,\gamma(y)$ is the duality relation for elements $g,\gamma\in\ell(Y)$, $\delta_y$ is the delta function at $y$, and $(\N^{\scriptscriptstyle\wedge}_t\circ\Phi_{\rm meas})^*:\ell(Y)\to\lh$ is the dual map of $\N^{\scriptscriptstyle\wedge}_t\circ\Phi_{\rm meas}$. If we set $\N'_t(y) = (\N^{\scriptscriptstyle\wedge}_t\circ\Phi_{\rm meas})^*(\delta_y)$, then the collection of measurements $(\N'_t)_{t\in T}$ so obtained is compatible, since such is the collection of classical measurements $(\N^{\scriptscriptstyle\wedge}_t)_{t\in T}$. Moreover, \eqref{eq:equ_proof} holds for $(\N'_t)_{t\in T}$, thus completing the proof.
\end{proof}

Remarkably, the converse statement of Theorem \ref{thm:classical} is not true. In other words, there exist state ensembles whose states do not commute, but for which the timing of partial information is irrelevant in specific guessing games. A nontrivial example is provided in Section \ref{sec:qubit_3}.

\section{Symmetry in guessing games}\label{sec:symmetry}

\subsection{ {Symmetries and group actions} 
}\label{sec:group_actions}

As we have seen in Theorem \ref{thm:reduction}, evaluating the maximal average score $\Efpost(\en)$ boils down to a usual state discrimination problem for the auxiliary state ensemble $\en_{f,\alpha}$ defined in \eqref{eq:enf-alpha}. 
However, finding the maximal guessing probability $\Pg(\en_{f,\alpha})$ may still be a difficult task since the number of states involved in the calculation scales as $|Y|^{|T|}$. Even assuming that the states of $\en$ are pure does not provide any actual simplification, as typically those of $\en_{f,\alpha}$ are mixed.

A natural attempt to reduce the complexity of the problem is by assuming that the state ensemble $\en$ possesses some symmetry, and then exploiting group theory in order to obtain the desired results. This indeed works for usual state discrimination \cite{Holevo73,ElMeVe04}, and our objective is now to provide an extension to the present more general setting.

For the rest of the section we fix a finite group $G$ acting on the sets $X$, $Y$ and $T$, and we assume that the quantities $f$ and $\alpha$ are $G$-invariant, i.e., invariant under the action of $G$ (see e.g. \cite{RFCG95} for the basics of group actions). 
More precisely, denoting by $g$ both an element of the group and its (left) action on the three sets above, we require that
\begin{enumerate}
\item[(S1)] $f(gx,gy) = f(x,y)$ for all $x\in X$, $y\in Y$ and $g\in G$,\label{it:cov1}
\item[(S2)] $\alpha(gt\mid gx) = \alpha(t\mid x)$ for all $x\in X$, $t\in T$ and $g\in G$.\label{it:cov2}
\end{enumerate}

In concrete situations, the above conditions often arise in a natural way. As  examples, we consider the cases of partition guessing games and deterministic posterior information described in Sections \ref{sec:partition} and \ref{sec:deterministic}. 
Thus, let $\upsilon:X\to Y$ be a surjective function and $(X_y)_{y\in Y}$ the partition of $X$ determined by $\upsilon$ as described in Section \ref{sec:partition}. Moreover, suppose the group $G$ acts on $X$ in a way that for all $y$ there is $y'$ such that $gX_y = \{gx : x\in X_y\} = X_{y'}$.
Then, we can define an action of $G$ on $Y$ by setting $gy = y'$. 
This action satisfies $\upsilon(gx)=g\upsilon(x)$ for all $x$ and $g$. Therefore, the score functions $f_\upsilon$ and $f_{\neg \upsilon}$ associated with $\upsilon$ are $G$-invariant (condition (S1)).
In the same way suppose $\tau:X\to T$ determines a partition $(X_t)_{t\in T}$ of $X$ which is preserved by the action of $G$. 
Then, the partial information map $\alpha_\tau$ of Section \ref{sec:non-overlapping} is invariant with respect to the action of $G$ on $T$ defined by $X_{gt} = gX_t$ (condition (S2)).

In order to describe symmetry on the operator side, we fix a projective unitary representation $U$ of $G$ on $\hh$ and we suppose that the state ensemble $\en$ is $G$-covariant in the following sense:
\begin{enumerate}
\item[(S3)] $U(g)\,\en(x)\,U(g)^* = \en(gx)$ for all $x\in X$ and $g\in G$.\label{it:cov3}
\end{enumerate}

\noindent{We can now state the following straightforward result.}

\begin{proposition}\label{prop:covariance}
If $f$, $\alpha$ and $\en$ satisfy conditions (S1)--(S3) above, then the auxiliary state ensemble $\en_{f,\alpha}$ defined in \eqref{eq:enf-alpha} satisfies 
\begin{equation}
U(g)\,\en_{f,\alpha}(\phi)\,U(g)^* = \en_{f,\alpha}(g.\phi)
\end{equation}
for all $\phi\in Y^T$ and $g\in G$, where the action of $G$ on $Y^T$ is defined as 
\begin{equation}
(g.\phi)(t) = g\phi(g^{-1}t)
\end{equation}
for all $t\in T$.
\end{proposition}
\noindent In other words, $G$-invariance of $f$ and $\alpha$ together with $G$-covariance of $\en$ imply $G$-covariance of $\en_{f,\alpha}$ if we regard the set $Y^T$ as a $G$-space in the natural way.

\subsection{ The case of an irreducible representation}

{If a guessing game possesses the symmetries described in the previous section, evaluating the maximal average score $\Efpost(\en)$ drastically simplifies if the representation $U$ is irreducible, i.e., $\{0\}$ and $\hh$ are the only subspaces of $\hh$ which are invariant under the action of $U$. Indeed, we have the following result.}

\begin{theorem}\label{thm:covariance}
Suppose $f$, $\alpha$ and $\en$ satisfy the symmetry conditions (S1)--(S3). Moreover, assume that the representation $U$ is irreducible. 
The following facts are true.
\begin{enumerate}[(a)]
\item Denote by $\Lambda(\en_{f,\alpha})$ the largest eigenvalue of all the operators $\en_{f,\alpha}(\phi)$, $\phi\in Y^T$. Then,
\begin{equation}
\Efpost(\en) = d\,|Y|^{|T|-1} \Delta(\en,f)\,\Lambda(\en_{f,\alpha}) \,.
\end{equation}
\item Fix $\phi_0\in Y^T$ such that the operator $\en_{f,\alpha}(\phi_0)$ has $\Lambda(\en_{f,\alpha})$ among its eigenvalues, and denote by $\Pi_0$ the orthogonal projection onto the eigenspace of $\en_{f,\alpha}(\phi_0)$ associated with $\Lambda(\en_{f,\alpha})$. 
The equality $\Efpost(\en;\bar{\M},\pi) = \Efpost(\en)$ is attained by the measurement\label{it:2_thm_covariance}
\begin{equation}\label{eq:M_opt_cov}
\bar{\M}(\phi) = \begin{cases}
\displaystyle d\,\big(\mo{G.\phi_0}\rank{\Pi_0}\big)^{-1}\, U(g)\,\Pi_0\,U(g)^* & \text{ if $\phi = g.\phi_0$ for some $g\in G$}\\
0 & \text{ otherwise}
\end{cases}\,.
\end{equation}
\end{enumerate}
\end{theorem}
\begin{proof}
By Proposition \ref{prop:covariance}, for all $g\in G$ we have
$$
\en_{f,\alpha}(g.\phi_0)\,U(g)\,\Pi_0\,U(g)^* = \Lambda(\en_{f,\alpha})\,U(g)\,\Pi_0\,U(g)^*\,.
$$
In particular, $\Pi_0$ commutes with $U(g)$ for all $g$ belonging to the stabilizer subgroup $G_0 = \{g\in G : g.\phi_0 = \phi_0\}$, and therefore the operator $\bar{\M}(\phi)$ given by \eqref{eq:M_opt_cov} is well defined. It also follows that $\en_{f,\alpha}(\phi)\,\bar{\M}(\phi) = \Lambda(\en_{f,\alpha})\,\bar{\M}(\phi)$ for all $\phi\in Y^T$. In order to apply Proposition \ref{prop:Pbound} to the state ensemble $\en_{f,\alpha}$ and the measurement $\bar{\M}$, we still need to check that $\sum_\phi\bar{\M}(\phi) = \id$. Indeed, since $U(g)\,\bar{\M}(\phi)\,U(g)^* = \bar{\M}(g.\phi)$ and
$$
U(g)\sum_\phi\bar{\M}(\phi)\,U(g)^* = \sum_\phi\bar{\M}(g.\phi) = \sum_\phi\bar{\M}(\phi) \,,
$$
Schur's lemma implies that $\sum_\phi\bar{\M}(\phi) = \mu\,\id$ for some $\mu\in\real$, where $\mu=1$ because
\begin{align*}
d\,\mu & = \tr{\mu\,\id} = \sum_\phi\tr{\bar{\M}(\phi)} = \sum_{\phi\in G.\phi_0} d\,|G.\phi_0|^{-1} = d \,.
\end{align*}
By Proposition \ref{prop:Pbound}, it then follows that $\Pg(\en_{f,\alpha}) = \Pg(\en_{f,\alpha};\bar{\M}) = d\,\Lambda(\en_{f,\alpha})$. Combining this fact with \eqref{eq:aux_Pgfpost} and \eqref{eq:reduction} yields the statement of the theorem.
\end{proof}

For all $\phi\in Y^T$, the set $G.\phi = \{g.\phi : g\in G\}$ is the orbit of $G$ passing through $\phi$. Item \eqref{it:2_thm_covariance} of the previous theorem means that we can always find an optimal measurement that is concentrated on such an orbit. As already remarked in the proof, the measurement \eqref{eq:M_opt_cov} satisfies the covariance condition
\begin{equation}
\bar{\M}(g.\phi) = U(g)\,\bar{\M}(\phi)\,U(g)^*
\end{equation}
for all $\phi\in Y^T$ and $g\in G$. This fact combined with the equality 
\begin{equation}
\pi_{gt}(gy\mid g.\phi) = \pi_t(y\mid\phi)
\end{equation}
implies that the marginals $(\N_t)_{t\in T}$ of $\bar{\M}$ are such that
\begin{equation}\label{eq:cov_Nt}
\N_{gt}(gy) = U(g)\,\N_t(y)\,U(g)^*
\end{equation}
for all $g\in G$, $y\in Y$ and $t\in T$. Therefore, different marginals are related by a permutation of the outcome set $Y$ and a unitary conjugation by $U$.

\section{Example: two pairs of orthogonal qubit states}\label{sec:qubit}

In the following we demonstrate the results of the previous sections by fixing four noncommuting qubit states as our state ensemble and evaluating $\Efpre(\en)$ and $\Efpost(\en)$ for several choices of $f$ and $\alpha$. 
In all the examples below, partial information increases the maximal average scores with both prior and posterior information. However, we will see two cases in which $\Efpre(\en) > \Efpost(\en)$ (Sections \ref{sec:qubit_1} and \ref{sec:qubit_2}) and one in which the timing of partial information is irrelevant (Section \ref{sec:qubit_3}).

\subsection{Notation}

We recall that the Hilbert space of a qubit system is $\hh=\C^2$ and that any qubit state $\varrho$ is represented as a vector in the Bloch ball $\{\vr\in\R^3 : \no{\vr}\leq 1\}$ by means of the relation
$$
\varrho = \tfrac{1}{2}\left(\id+\vr\cdot\vsigma\right)\,.
$$
In this formula we have denoted $\vr\cdot\vsigma = r_1\sigma_1 + r_2\sigma_2 + r_3\sigma_3$ for the vector $\vr=r_1\ve_1+r_2\ve_2+r_3\ve_3$, where $\sigma_1$, $\sigma_2$ and $\sigma_3$ are the three Pauli matrices and $\ve_1$, $\ve_2$ and $\ve_3$ the unit vectors along the coordinate axes. More generally, any selfadjoint operator $M\in\lc$ can be written as
$$
M = \mu\,\id + \vm\cdot\vsigma
$$
for some $\mu\in\R$ and $\vm\in\R^3$, uniquely detemined by $M$.
If $\vm$ is nonzero, the eigenvalues $\lambda_+$, $\lambda_-$ of $M$ and the corresponding eigenprojections $\Pi_+$, $\Pi_-$ are
\begin{align*}
\lambda_\pm = \mu\pm\no{\vm}\,,\qquad\qquad \Pi_\pm = \tfrac{1}{2}\left(\id\pm\vmh\cdot\vsigma \right) \, ,
\end{align*}
where $\vmh = \vm/\no{\vm} = \vm/(\lambda_+ - \mu)$ is the unit vector along the direction of $\vm$.

\begin{figure}[h!]
\centering\includegraphics{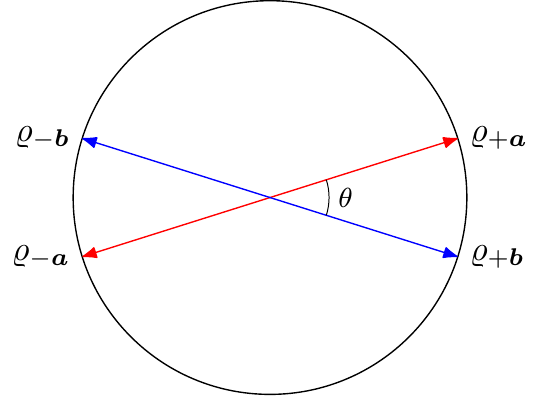}
\caption{The states of the ensemble \eqref{eq:4states} represented in a section of the Bloch ball. Each state is chosen with uniform probability and is directed along one of the labels $+\va$, $-\va$, $+\vb$ and $-\vb$.\label{fig:Bloch}}
\end{figure}

For $\theta\in (0,\pi/2]$, we fix
\begin{equation}
\va = \cos\left(\half\theta\right) \ve_1 + \sin\left(\half\theta\right) \ve_2\,,\qquad\qquad \vb = \cos\left(\half\theta\right) \ve_1 - \sin\left(\half\theta\right) \ve_2
\end{equation}
and define
$$
X = \{+\va,\,-\va,\,+\vb,\,-\vb\}
$$
as the label set of Alice. 
The state ensemble $\en$ is chosen to be
\begin{equation}\label{eq:4states}
\en(\vx) = \tfrac{1}{8}\left(\id+\vx\cdot\vsigma\right)
\end{equation}
for all $\vx\in X$.
It hence corresponds to two orthogonal pairs of pure states, $\varrho_{+\va}$, $\varrho_{-\va}$ and $\varrho_{+\vb}$, $\varrho_{-\vb}$, all apprearing with the same probability $1/4$ in the state ensemble $\en$ (see Fig. \ref{fig:Bloch} for an illustration in the Bloch ball).

The elements of $X$ are permuted by the dihedral group $D_2\subset SO(3)$, which consists of the identity element $I$ together with the three $180^\circ$ rotations $R_1$, $R_2$ and $R_3$ along the respective coordinate axes. The group $D_2$ acts on $\C^2$ by means of the projective unitary representation
\begin{equation}
U(I) = \id \, , \qquad\qquad U(R_i) = \sigma_i \,,
\end{equation}
and the state ensemble $\en$ is manifestly $D_2$-covariant. Since the representation $U$ is irreducible, we can use Theorem \ref{thm:covariance} to evaluate $\Efpost(\en)$ provided that $f$ and $\alpha$ are $D_2$-invariant.

\subsection{Discrimination and antidiscrimination without partial information}\label{sec:qubit_0}

For comparison we recall the maximal guessing probabilities in the usual discrimination and antidiscrimination guessing games when there is no partial information available.
By using Proposition \ref{prop:Pbound}, it is straightforward to show that $\Pg(\en)=1/2$ irrespective of the angle $\theta$. 
A different proof of this fact can be found e.g.~in \cite{Bae13}. 

The maximal guessing probability in the antidiscrimination guessing game is $1$, as it can be evaluated by forming first the auxiliary state ensemble given in \eqref{eq:enf_0}.
An alternative way to see this fact is by observing that $\sum_{\vx\in X} \varrho_\vx = 2\,\id$. 
This condition implies that the four states can be perfectly antidiscriminated with any prior probability distribution $p$ \cite{HeKe18}.

\subsection{Discrimination with deterministic posterior information}\label{sec:qubit_1}

Let us consider discrimination of the state ensemble \eqref{eq:4states} with deterministic posterior information, hence we choose $X=Y$ and $f=f_\delta$. 
The set $X$ is partitioned into two disjoint subsets $X_a$ and $X_b$, where
\begin{equation}\label{eq:qubit_partition}
X_a = \{+\va,\,-\va\}\,,\qquad\qquad X_b = \{+\vb,\,-\vb\}\,.
\end{equation}
The partial information consists in giving the correct subset of the input label, thus $T=\{a,b\}$ and the partial information map is $\alpha_\tau$ with $\tau(\pm\va) = a$ and $\tau(\pm\vb) = b$ (see Section \ref{sec:non-overlapping}).

We begin by evaluating the maximal average score $\Epre_{f_\delta,\alpha_\tau}(\en)$. 
It is enough to observe that the conditional state ensemble \eqref{eq:ent} is
\begin{equation}
\en_t(\vx) = \begin{cases}
\frac{1}{4} \left(\id+\vx\cdot\vsigma\right) & \text{ if $\vx\in X_t$}\\
0 & \text{ otherwise}
\end{cases}
\end{equation}
and that $\en_t$ is perfectly discriminated by means of the sharp measurement
\begin{equation}\label{eq:4states-ex1-opti-meas}
\N_t(\vx) = \begin{cases}
\frac{1}{2}\left(\id+\vx\cdot\vsigma\right) & \text{ if $\vx\in X_t$}\\
0 & \text{ otherwise}
\end{cases} \,.
\end{equation}
It follows that
\begin{equation}\label{eq:4states-ex1-pre}
\Epre_{f_\delta,\alpha_\tau}(\en) = 1 \,.
\end{equation}

In order to calculate the maximal average score $\Epost_{f_\delta,\alpha_\tau}(\en)$ in the posterior information guessing game, we use the symmetry of the problem.
The score function $f_\delta$ and the partial information map $\alpha_\tau$ are $D_2$-invariant by the discussion {after conditions (S1) and (S2) in Section \ref{sec:group_actions}.} Thus, by Theorem \ref{thm:covariance}, evaluating the maximal average score $\Epost_{f_\delta,\alpha_\tau}(\en)$ amounts to finding the maximal eigenvalue of the operators $\en_{f_\delta,\alpha_\tau}(\phi)$, $\phi\in Y^T$, defined by \eqref{eq:enfalpha_discr2}, which in the current case become
\begin{equation}
\en_{f_\delta,\alpha_\tau}(\vx_1,\vx_2) = \frac{C}{8}\cdot\begin{cases}
\left(\id + \vx_1\cdot\vsigma\right) & \text{ if $\vx_1,\vx_2\in X_1$}\\
\left(\id + \vx_2\cdot\vsigma\right) & \text{ if $\vx_1,\vx_2\in X_2$}\\
\left[2\,\id + \left(\vx_1 + \vx_2\right)\cdot\vsigma\right] & \text{ if $\vx_1\in X_1$ and $\vx_2\in X_2$} \\
0 & \text{ if $\vx_1\in X_2$ and $\vx_2\in X_1$}
\end{cases}
\end{equation}
with $1/C=|Y|^{|T|-1}\Delta(\en,d)=4$.
By means of straightforward calculations, we get
$$
\Lambda (\en_{f_\delta,\alpha_\tau})= \frac{C}{8}\,\big(2+\big\|\va+\vb\big\|\big) = \frac{C}{4} \left(1+\sqrt{\frac{1+\cos\theta}{2}}\right) \,,
$$
and then Theorem \ref{thm:covariance} yields
\begin{equation}\label{eq:4states-ex1-post}
\Epost_{f_\delta,\alpha_\tau}(\en) = \frac{1}{2} \left(1+\sqrt{\frac{1+\cos\theta}{2}}\right)\,.
\end{equation}

The average scores \eqref{eq:4states-ex1-pre} and \eqref{eq:4states-ex1-post} were already obtained in \cite{CaHeTo18}, where a detailed description of the optimal measurements was also provided. 
We remark that there are strict inequalities
\begin{equation}
\Epre_{f_\delta,\alpha_\tau}(\en) > \Epost_{f_\delta,\alpha_\tau}(\en) > \Eg_{f_\delta}(\en)
\end{equation}
for all $\theta\in (0,\pi/2]$ and of these three quantities only $\Epost_{f_\delta,\alpha_\tau}(\en)$ varies with $\theta$.

\subsection{Discrimination by excluding one wrong option}\label{sec:qubit_2}

Let us still consider the discrimination game, but now with a different kind of partial information. Namely, Alice excludes one wrong option.
We hence keep $X=Y$ and $f=f_\delta$, but now $X=T$ and the partial information map is $\alpha_{\rm ex}$ described in Section \ref{sec:non-deterministic}, that is, $\alpha_{\rm ex}(\vt\mid\vx) = \left(1-\delta_{\vx,\vt}\right)/3$.

In the present case, the conditional state ensemble \eqref{eq:ent} is
\begin{equation}\label{eq:4states-ex2-ent}
\en_\vt(\vx) = \tfrac{1}{6}\left(1-\delta_{\vx,\vt}\right)\left(\id+\vx\cdot\vsigma\right) \,.
\end{equation}
Using Proposition \ref{prop:Pbound} we conclude that $\Eg_{f_\delta}(\en_\vt) = \Pg(\en_\vt) = 2/3$, the unique optimal measurement being still given by \eqref{eq:4states-ex1-opti-meas} with $t=\tau(\vt)$. 
The maximal average score with prior information is then
\begin{equation}
\Epre_{f_\delta,\alpha_{\rm ex}}(\en) = \frac{2}{3} \,.
\end{equation}
Since the sharp optimal measurements \eqref{eq:4states-ex1-opti-meas} do not commute for $t\neq t'$, we expect that $\Epre_{f_\delta,\alpha_{\rm ex}}(\en) > \Epost_{f_\delta,\alpha_{\rm ex}}(\en)$.

With the introduced group theoretical machinery we can find out that
\begin{equation}\label{eq:qubit-wrong-post}
\Epost_{f_\delta,\alpha_{\rm ex}}(\en) = \frac{1}{12} \left(4+\sqrt{10+6\cos\theta}\right)\,.
\end{equation}
To see this, we first observe that the partial information map $\alpha_{\rm ex}$ is $D_2$-invariant, hence Theorem \ref{thm:covariance} applies also in this case. The auxiliary state ensemble \eqref{eq:ruling-out_enf} becomes
\begin{equation}
\begin{aligned}
& \en_{f_\delta,\alpha_{\rm ex}}(\phi) = \frac{C}{24} \sum_{\vx} \mo{\phi^{-1}(\vx)\setminus\{\vx\}} \left(\id+\vx\cdot\vsigma\right) \\
& \qquad = \frac{C}{24}\, \Big\{ \big(\alpha^\phi_+ + \alpha^\phi_- + \beta^\phi_+ + \beta^\phi_-\big)\,\id + \Big[\big(\alpha^\phi_+ - \alpha^\phi_-\big)\,\va + \big(\beta^\phi_+ - \beta^\phi_-\big)\,\vb\Big]\cdot\vsigma\Big\}\,,
\end{aligned}
\end{equation}
where $1/C=|Y|^{|T|-1}\Delta(\en,d)=64$ and we have denoted
\begin{equation}
\alpha^\phi_\pm = \mo{\phi^{-1}(\pm\va)\setminus\{\pm\va\}}\,,\qquad\qquad \beta^\phi_\pm = \big|\phi^{-1}(\pm\vb)\setminus\{\pm\vb\}\big|\,.
\end{equation}
The largest eigenvalue of $\en_{f_\delta,\alpha_{\rm ex}}(\phi)$ is
\begin{equation*}
\begin{aligned}
\lambda(\phi) & = \frac{C}{24}\, \Big\{\alpha^\phi_+ + \alpha^\phi_- + \beta^\phi_+ + \beta^\phi_- + \Big\|\big(\alpha^\phi_+ - \alpha^\phi_-\big)\,\va + \big(\beta^\phi_+ - \beta^\phi_-\big)\,\vb\Big\|\Big\} \\
& = \frac{C}{24}\,\gamma\big(\alpha^\phi_+,\,\alpha^\phi_-,\,\beta^\phi_+,\,\beta^\phi_-\big)\,,
\end{aligned}
\end{equation*}
where $\gamma$ is the function
\begin{equation*}
\begin{aligned}
& \gamma\big(\alpha_+,\,\alpha_-,\,\beta_+,\,\beta_-\big) = \alpha^\phi_+ + \alpha^\phi_- + \beta^\phi_+ + \beta^\phi_- \\
& \qquad\qquad + \big[\big(\alpha_+ - \alpha_-\big)^2 + \big(\beta_+ - \beta_-\big)^2 + 2\,\big(\alpha_+ - \alpha_-\big)\big(\beta_+ - \beta_-\big)\cos\theta\big]^{\frac{1}{2}} \,.
\end{aligned}
\end{equation*}
The corresponding eigenprojection is
$$
\Pi(\phi) = \tfrac{1}{2}\left(\id+\vmh(\phi)\cdot\vsigma\right)
$$
with
$$
\vmh(\phi) = \frac{\big(\alpha^\phi_+ - \alpha^\phi_-\big)\,\va + \big(\beta^\phi_+ - \beta^\phi_-\big)\,\vb}{\gamma\big(\alpha_+,\,\alpha_-,\,\beta_+,\,\beta_-\big) - \big(\alpha^\phi_+ + \alpha^\phi_- + \beta^\phi_+ + \beta^\phi_-\big)}\,.
$$
For all $\phi\in X^X$, the numbers $\alpha^\phi_\pm$ and $\beta^\phi_\pm$ satisfy the constraints
\begin{equation*}
\alpha^\phi_\pm,\beta^\phi_\pm\in\naturale\,,\qquad\quad \alpha^\phi_\pm,\beta^\phi_\pm \leq |X|-1\,,\qquad\quad \alpha^\phi_+ + \alpha^\phi_- + \beta^\phi_+ + \beta^\phi_- \leq |X|\,.
\end{equation*}
The maximum of $\gamma\big(\alpha_+,\,\alpha_-,\,\beta_+,\,\beta_-\big)$ with $\alpha_\pm$, $\beta_\pm$ subject to these constraints is equal to $4+\sqrt{10+6\cos\theta}$ (see Appendix \ref{app:max} for details) and it is attained at the feasible points
$$
f_0 = (1,0,3,0),\,\qquad f_1 = (3,0,1,0),\,\qquad f_2 = (0,3,0,1),\,\qquad f_3 = (0,1,0,3) \,.
$$
If $\phi_0\in X^X$ is given by
\begin{equation}
\phi_0(+\va) = \phi_0(-\va) = \phi_0(-\vb) = +\vb\,,\qquad\qquad \phi_0(+\vb) = +\va
\end{equation}
and we further define
\begin{equation}
\phi_i = R_i.\phi_0 \quad\text{for } i=1,2,3\,,
\end{equation}
then with straightforward calculations
$$
f_i = \big(\alpha^{\phi_i}_+,\,\alpha^{\phi_i}_-,\,\beta^{\phi_i}_+,\,\beta^{\phi_i}_-\big) \quad\text{for all } i=0,1,2,3\,.
$$
Using the notations of Theorem \ref{thm:covariance}, it follows that
$$
\Lambda(\en_{f_\delta,\alpha_{\rm ex}}) = \frac{C}{24}\,\left(4+\sqrt{10+6\cos\theta}\right)
$$
and the operator $\en_{f_\delta,\alpha_{\rm ex}}(\phi_0)$ has $\Lambda(\en_{d,\alpha_{\rm ex}})$ among its eigenvalues. Therefore, we obtain \eqref{eq:qubit-wrong-post}.

The optimal measurement \eqref{eq:M_opt_cov} is
\begin{equation}
\bar{\M}(\phi) = \begin{cases}
\tfrac{1}{4}\left(\id+\vmh(\phi)\cdot\vsigma\right) & \text{ if $\phi\in\{\phi_0,\phi_1,\phi_2,\phi_3\}$} \\
0 & \text{ otherwise}
\end{cases}
\end{equation}
with
\begin{equation}
\begin{aligned}
\vmh(\phi_0) & = - \vmh(\phi_2) = \frac{\va+3\vb}{\sqrt{10+6\cos\theta}}\,, \\
\vmh(\phi_1) & = - \vmh(\phi_3) = \frac{3\va+\vb}{\sqrt{10+6\cos\theta}}\,.
\end{aligned}
\end{equation}
Its marginal $\N_{+\va}$ is
\begin{align*}
\N_{+\va}(+\va) & = 0\,, & \quad \N_{+\va}(+\vb) & = \tfrac{1}{4}\left[2\,\id+\left(\vmh(\phi_0)+\vmh(\phi_1)\right)\cdot\vsigma\right],\\
 \N_{+\va}(-\va) & = \tfrac{1}{4}\left(\id-\vmh(\phi_1)\cdot\vsigma\right), & \quad \N_{+\va}(-\vb) & = \tfrac{1}{4}\left(\id-\vmh(\phi_0)\cdot\vsigma\right),
\end{align*}
and the other marginals $\N_{-\va}$, $\N_{+\vb}$ and $\N_{-\vb}$ are obtained from $\N_{+\va}$ by means of the relation \eqref{eq:cov_Nt}.

\subsection{Partition guessing game by excluding one wrong option}\label{sec:qubit_3}

Finally, we consider a partition guessing game of the kind described in Section \ref{sec:partition}.
We choose $Y = \{a,b\}$ and let $\upsilon:X\to Y$ be the function $\upsilon(\pm\va)=a$, $\upsilon(\pm\vb)=b$. With this choice of $Y$ and $\upsilon$, we consider the score function $f_\upsilon$ defined in \eqref{eq:partition}. 
Thus, the task is to detect the correct direction of the label $\vx$, i.e., to guess whether $\vx\in X_a$ or $\vx\in X_b$ for the two sets $X_a$, $X_b$ defined in \eqref{eq:qubit_partition}. 
We still have $X=T$ and the partial information map is $\alpha_{\rm ex}(\vt\mid \vx) = (1-\delta_{\vx,\vt})/3$ as in the previous section.

We first observe that, according to Section \ref{sec:partition}, without partial information the task is equivalent to discriminating two totally mixed states. Indeed, in the current case, $\en_{f_\upsilon}(y) = (1/4)\,\id$ for both $y=a,b$, and therefore the best discrimination strategy is random guessing, i.e., 
\begin{equation}
\Eg_{f_\upsilon} (\en) = \Pg(\en_{f_\upsilon}) = \frac{1}{2} \, .
\end{equation}
In other words, we can reach the maximal average score without making any measurement.

To calculate the optimal average score in the cases with partial information, we first observe that the conditional state ensemble $\en_\vt$ is the same as \eqref{eq:4states-ex2-ent}, but now the score function has changed. We evaluate $\Eg_{f_\upsilon}(\en_\vt)$ by using \eqref{eq:aux_Pgfpost_1}-\eqref{eq:aux_Pgfpost_2}, where in the present case $\Delta(\en,f_\upsilon)=1$ and the auxiliary state ensemble \eqref{eq:enf_0} is
$$
(\en_\vt)_{f_\upsilon} (y) = \frac{1}{6}\cdot\begin{cases}
\left(\id -\vt\cdot\vsigma\right) & \text{ if $y=\upsilon(\vt)$}\\
2\,\id & \text{ otherwise}
\end{cases}\,.
$$
We obtain
$$
\Eg_{f_\upsilon}(\en_\vt) = \Delta(\en,f_\upsilon)\ \Pg\big((\en_\vt)_{f_\upsilon}\big) = \frac{2}{3} \,,
$$
where we used Proposition \ref{prop:Pbound} to evaluate $\Pg\big((\en_\vt)_{f_\upsilon}\big) = 2/3$. A measurement $\N_\vt$ maximizing $\Eg_{f_\upsilon}(\en_\vt;\N_\vt) = \Delta(\en,f_\upsilon)\ \Pg\big((\en_\vt)_{f_\upsilon};\N_\vt\big)$ is the trivial measurement
\begin{equation}
\N_\vt(y) = \left(1-\delta_{y,\upsilon(\vt)}\right)\id\,.
\end{equation}
Clearly, the collection of measurements $(\N_\vt)_{\vt\in T}$ is compatible. By \eqref{eq:sup_any} and \eqref{eq:sup_compatible}, it follows that
\begin{equation}
\Epre_{f_\upsilon,\alpha_{\rm ex}} (\en) = \Epost_{f_\upsilon,\alpha_{\rm ex}} (\en) = \frac{2}{3}
\end{equation}
independently of the angle $\theta$.
As in the earlier consideration of the same task but without partial information, also in this case the maximal average score can be reached without making any measurement.

\section{Conclusion and outlook}\label{sec:conc}

In minimum error state discrimination the task is to correctly guess the unknown state of a quantum system from a finite set of alternatives. Guessing games constitute a natural extension to tasks that do not necessarily require the full determination of the unknown state. The difference between the two scenarios is in the choice of the figure of merit, which is assumed to be Kronecker delta for state discrimination, and is allowed to be any score function (even with possibly different input and output sets) for guessing games. Regarded in this way, the history of guessing games traces back to the very origin of state discrimination, since in their seminal works Holevo \cite{Holevo73} and Helstrom \cite{QDET76} already considered a figure of merit of a general type. Within this well-established framework, our contribution was a systematic study of the role of partial information as a resource for improving the score of the games. Actually, it is posterior information that fundamentally changes the usual scenario and makes the already known optimization techniques for state discrimination not directly exportable to the new context. Nevertheless, we showed that even in this case all earlier results become applicable at the cost of switching from the original game to a properly derived auxiliary state discrimination task.

There are several interesting generalizations of guessing games with partial information beyond the scenarios described in this paper. Firstly, in our approach we optimized the game only on Bob's side, i.e., on the side which receives information and tries to retrieve the original message encoded by the sender.
{However,} also Alice could try to arrange her preparation in order to improve the score of the game. In the scenario without partial information, this amounts to maximizing the average score \eqref{eq:Ef} both over the measurement $\M$ and the state ensemble $\en$. If partial information is taken into account, then also the partial information map $\alpha$ can enter the optimization problem for the average score \eqref{eq:Pgfpost}. When Alice is allowed to cooperate in the game, {however,} suitable constraints should be imposed over the encodings that she can access{, as otherwise the game becomes trivial. Indeed, without any constraint, Alice can always use the state ensemble $\en(x) = \delta_{x_0,x}\,\varrho$, and then Bob's optimal strategy is guessing any $y$ which maximizes $f(x_0,y)$, with no reference to Alice's partial information $t$ and his measurement outcome $z$.} For example, the prior probability distribution $p(x) = \tr{\en(x)}$ may be required to be uniform, a constraint that is frequent in communication protocols.

As a second possible generalization, one may consider the case in which also posterior information is of the quantum type, i.e., the partial information map $\alpha$ is a collection of quantum states $(\alpha_x)_{x\in X}$. In this case, the average score \eqref{eq:Pgfpost} becomes
\begin{equation}\label{eq:Pgfpost_noncomm}
\Efpost(\en;\M,\Oo) =  \sum_{x,y,z} f(x,y)\, \tr{\alpha_x\,\Oo_z(y)}\,\tr{\en(x)\,\M(z)} \,,
\end{equation}
where each $\Oo_z$ is a measurement with the outcome set $Y$ and the correspondence $\Oo : Z\mapsto \Oo_z$ takes the place of the post-processing map $\nu$. Indeed, {in the classical case} all states $\alpha_x$, $x\in X$, are diagonal with respect to the same basis $(\varphi_t)_{t\in T}$, and then \eqref{eq:Pgfpost_noncomm} boils down to \eqref{eq:Pgfpost} if one replaces $\alpha(t\mid x) = \tr{\kb{\varphi_t}{\varphi_t}\,\alpha_x}$ and $\nu_t(y\mid z) = \tr{\kb{\varphi_t}{\varphi_t}\,\Oo_z(y)}$. In the latter replacement, the measurement $\Oo_z$ just consists in reading out the classical message $t$ and, according to its value and the value of $z$, it yields the outcome $y$ with probability $\nu_t(y\mid z)$. 
Although extending \eqref{eq:Pgfpost} to quantum posterior information, the average score \eqref{eq:Pgfpost_noncomm} has a different statistical interpretation. 
Namely, formula \eqref{eq:Pgfpost_noncomm} describes a scenario in which Bob receives two quantum states at a time, that is, $\alpha_x$ and $\varrho_x = \en(x)/\tr{\en(x)}$. He then performs the measurement $\M$ on $\varrho_x$ and, according to the result $z$ of this measurement, he makes a successive measurement $\Oo_z$ on $\alpha_x$. Finally, based on the outcome $y$ of the measurement $\Oo_z$, Bob gets the score $f(x,y)$. As we see, in this scenario partial information is still used to post-process the measurement $\M$, but in a way that depends on the result of another measurement. 
Of course, the same alternative interpretation is also valid for the average score \eqref{eq:Pgfpost} if we regard $t$ as the conditioning variable in the probabilities $\nu_t(y\mid z)$. 
Interestingly, the two {interpretations} coexist when partial information is of classical type, as they merely differ in the order of conditioning over $z$ and $t$. However, only one of them makes sense when partial information is turned into quantum, since conditioning over $t$ is then no longer possible.

As we illustrated in the paper, one of the main applications of guessing games with partial information is the detection of quantum incompatibility. In Section \ref{sec:incomp}, we already pointed out several related questions which still remain unsolved. The central one is characterizing guessing games that are capable of detecting all the incompatible collections of measurements with a given length. We showed that, if $|T|$ is the length and $Y$ is the outcome set of the measurements, then incompatibility can always be detected by using a state ensemble of $|Y|\,|T|$ states. However, it is not clear whether smaller state ensembles still suffice for the task. An interesting related question is to characterize those guessing games which do not detect incompatibility at all, i.e., for which games the timing of partial information is irrelevant. Although we proved that this is always the case for commutative state ensembles, we also showed in Section \ref{sec:qubit_3} that commutativity is not a necessary condition, and other features of the game (i.e., its score function and/or partial information map) must come into play.

\newpage

\appendix

\section{Auxiliary calculations for Section \ref{sec:qubit}}\label{app:max}

For the function
\begin{equation*}
\begin{aligned}
& \gamma\big(\alpha_+,\,\alpha_-,\,\beta_+,\,\beta_-\big) = \alpha_+ + \alpha_- + \beta_+ + \beta_- \\
& \qquad\qquad + \big[\big(\alpha_+ - \alpha_-\big)^2 + \big(\beta_+ - \beta_-\big)^2 + 2\,\big(\alpha_+ - \alpha_-\big)\big(\beta_+ - \beta_-\big)\cos\theta\big]^{\frac{1}{2}}
\end{aligned}
\end{equation*}
with $\theta\in (0,\pi/2]$, we evaluate
\begin{gather}
\max\gamma\big(\alpha_+,\,\alpha_-,\,\beta_+,\,\beta_-\big)\quad\text{subject to}\nonumber \\
\alpha_\pm,\beta_\pm\in\{0,1,2,3\}\quad\text{and}\quad \alpha_+ + \alpha_- + \beta_+ + \beta_- \leq 4 \label{eq:constraint1_bis}
\end{gather}
and we also find the set of feasible points where the maximum is attained.
Assuming \eqref{eq:constraint1_bis}, we distinguish three cases.
\begin{enumerate}[(i)]
\item Suppose $\alpha_+ = \alpha_- = 0$. Then,
\begin{align*}
\gamma\big(\alpha_+,\,\alpha_-,\,\beta_+,\,\beta_-\big) = \beta_+ + \beta_- + \mo{\beta_+ - \beta_-} = 2\max\{\beta_+,\,\beta_-\} \leq 6\,.
\end{align*}
\item Suppose $\beta_+ = \beta_- = 0$. Then, $\gamma\big(\alpha_+,\,\alpha_-,\,\beta_+,\,\beta_-\big)\leq 6$ as above.
\item Suppose $\alpha_j\neq 0$ and $\beta_k\neq 0$ for some $j,k$. We have
\begin{gather*}
\big(\alpha_+ + \alpha_-\big)^2 \geq \big(\alpha_+ - \alpha_-\big)^2\,,\qquad\qquad \big(\beta_+ + \beta_-\big)^2 \geq \big(\beta_+ - \beta_-\big)^2 \\
\big(\alpha_+ + \alpha_-\big) \big(\beta_+ + \beta_-\big) \geq \big(\alpha_+ - \alpha_-\big) \big(\beta_+ - \beta_-\big) \,.
\end{gather*}
Moreover, the equality is attained in all the three relations if and only if either $\alpha_+ = \beta_+ = 0$ or $\alpha_- = \beta_- = 0$. Then,
\begin{align*}
\gamma\big(\alpha_+,\,\alpha_-,\,\beta_+,\,\beta_-\big) & \leq \gamma\big(\alpha_+ + \alpha_-,\,0,\,\beta_+ + \beta_-,\,0\big) \\
& \leq \gamma\big(\alpha_+ + \alpha_-,\,0,\,4 - \big(\alpha_+ + \alpha_-\big),\,0\big) \,,
\end{align*}
where $\alpha_+ + \alpha_- \geq 1$ and $4 - \big(\alpha_+ + \alpha_-\big) \geq \beta_+ + \beta_- \geq 1$. In the last expression, both relations are equalities if and only if $\beta_+ + \beta_- = 4 - \big(\alpha_+ + \alpha_-\big)$ and either $\alpha_+ = \beta_+ = 0$ or $\alpha_- = \beta_- = 0$. It is easy to see that
\begin{align*}
\max\{\gamma(\alpha,\,0,\,4 - \alpha,\,0) : \alpha\in\{1,2,3\}\} & = 4+\sqrt{10+6\cos\theta}
\end{align*}
and that the maximum is attained if and only if $\alpha\in\{1,3\}$. Since
$$
\gamma\big(\alpha_+ + \alpha_-,\,0,\,\beta_+ + \beta_-,\,0\big) = \gamma\big(0,\,\alpha_+ + \alpha_-,\,0,\,\beta_+ + \beta_-\big) \,,
$$
we conclude that $\gamma\big(\alpha_+,\,\alpha_-,\,\beta_+,\,\beta_-\big) = 4+\sqrt{10+6\cos\theta}$ if and only if the quadruple $\big(\alpha_+,\,\alpha_-,\,\beta_+,\,\beta_-\big)$ belongs to the set
\begin{equation*}
F=\{(1,0,3,0),\,(3,0,1,0),\,(0,1,0,3),\,(0,3,0,1)\}
\end{equation*}
\end{enumerate}
Combining the three cases above, we see that the constrained maximum of $\gamma$ is $4+\sqrt{10+6\cos\theta}$, and that the set $F$ constitutes all the feasible points at which the maximum is attained.

\end{document}